\documentclass[a4paper,article,10pt]{memoir}
\usepackage[english]{babel}
\usepackage{CCLAuthor}
\usepackage[reqno]{amsmath}
\usepackage{amssymb}
\usepackage{amsthm}  
\theoremstyle{plain}

\theoremstyle{definition}

\numberwithin{theorem}{chapter}
\usepackage[T1]{fontenc}
 \usepackage{epsfig} 
\usepackage{slashbox}
\usepackage[round,authoryear]{natbib} 
\usepackage[]{graphicx}               
\usepackage{amsmath, amsfonts, amsthm, amssymb, amstext}
 \usepackage{mathrsfs}
  \usepackage{ucs}
\newcommand{\ds}{\displaystyle}
\newcommand{\ve}{\varepsilon}

\begin{document}

%
%
%
\title{Asymptotic Analysis and Synthesis 
in Mechanics of Solids and Nonlinear Dynamics}
%
%
\author{%
    I.V. Andrianov\textsuperscript{*}, H. Topol\textsuperscript{*}
    \\ \smallskip\small
    \textsuperscript{*} 
    Institute of General Mechanics, 
    RWTH Aachen University, D-52062 Aachen , Germany
    }
%
%
    \maketitle
%

%
%

    \noindent \begin{abstract}
 In this lectures  various methods which give a possibility to extend an area of applicability of perturbation series and hence to omit their local character are analysed. 
While applying asymptotic methods as a rule the following situation appears: the existence of asymptotics for $\ve\rightarrow 0$  implies an existence of the asymptotics for 
$\ve\rightarrow\infty$. Therefore, the 
idea to construct one function valid for the whole parameter interval for $\ve$ is very attractive. The construction of asymptotically equivalent functions possessing  a
 known asymptotic behaviour for $\ve\rightarrow 0$ and $\ve\rightarrow \infty$ will be discussed. 
Using summation and interpolation procedures we focus on continuous models derived from a discrete micro-structure. Various 
continualization procedures that take the non-local interaction between variables of the discrete media into account are analysed. 
    \end{abstract}

\tableofcontents
\CCLsection{Overview of Manuscript Preparation and Delivery}

Asymptotic analysis is a constantly growing branch of mathematics which influences the development of various pure and applied sciences. 
The famous mathematicians  \citet{book31} and \citet{book69} said that an asymptotic description is not only a suitable instrument for 
the mathematical analysis of nature but that it also has an additional deeper intrinsic meaning, and that the asymptotic approach is more than
 just a mathematical technique; it plays a rather fundamental role in science. 
And here it appears that the many existing asymptotic methods 
comprise a set of approaches that in some way belong rather to art than to science. \citet{book44} even introduced special term ``asymptotology''
and defined it as the ``art of dealing with applied mathematical systems in limiting cases``.
Here it should be noted that he called for a formalization of the accumulated experience to convert the art of asymptotology 
into a science of asymptotology. \\
      Asymptotic methods for solving mechanical and physical problems have been developed by many authors. 
We can mentioned the excellent monographs by \cite{book38}, \cite{book43}, \cite{book60}, \cite{book62, book63}, 
 \cite{book80, book81}, \cite{book83}, and many others. The main feature of the present book can be formulated as follows: 
it deals with new trends and applications of asymptotic approaches in the fields of Nonlinear Mechanics and Mechanics of Solids. 
It illuminates the developments in the field of asymptotic mathematics from different viewpoints, reflecting the field's multidisciplinary nature. 
The choice of topics reflects the authors' own research experience and participation in applications. The authors have paid special
 attention to examples and discussions of results, and have tried to avoid burying the central ideas in formalism, notations, and technical details.

\CCLsection{Some Nonstandard Perturbation Procedures}
\CCLsubsection{Choice of Small Parameters}
The choice of the asymptotic method and the introduction of small dimensionless parameters in the system is very often 
the most significant and informal part of the analytical study of physical problems. 
This should help the experience and intuition, analysis of the physical nature of the problem, experimental and numerical results.  
It often dictated by physical considerations, is clearly shown as dimensionless and scaling. 
However, it is sometimes advantageous to use is not obvious, and perhaps even strange at first glance, 
the initial approximation. To illustrate this, consider a simple example \citep{book17}: algebraic equation 
\begin{equation}\label{eq1}
 x^5+x=1.
\end{equation}
We seek the real root of Eq. (\ref{eq1}), the exact value of which can be determined numerically: $x=0.75487767\dots\;$.
A small parameter $\ve$ is not included explicitly in Eq. (\ref{eq1}). Consider the various possibilities of introducing a parameter $\ve$ in Eq. (\ref{eq1}).
\begin{enumerate}
\item We introduce a small parameter $\ve$ with a nonlinear term of Eq. (\ref{eq1})
\begin{equation}\label{eq2}
 \ve x^5+x=1,
\end{equation}
and present $x$  as a series of $\ve$ 
\begin{equation}\label{eq3}
 x=a_0+a_1\ve+a_2\ve^2+\dots.
\end{equation}
Substituting the series (\ref{eq3}) into Eq. (\ref{eq2}) and equating terms of equal powers, we obtain\\\\
$
 a_0=1,\quad a_1=-1,\quad a_2=5,\quad a_3=-35,\quad a_4=285,\quad a_5=-2530,\\ a_6=23751.
$\\\\
These values can be obtained by a closed expression for the coefficients $a_n$:
\[
 a_n=\dfrac{(-1)^n(5n)!}{n!(4n+1)!}.
\]
   
The radius $R$ of convergence of the series (\ref{eq3}) is $R=\frac{4^4}{5^5}=0.08192$.
Consequently, for $\ve=1$ series (\ref{eq3}) diverges very fast, so the sum of the first six terms is 21476. 
The situation can be corrected by the method of Pad\'{e} approximants (see Sect. 2). 
Constructing a Pad\'{e} approximant with three terms in the numerator and denominator and calculating it with  $\ve=1$, 
we obtain the value of the root $x = 0.76369$ (the deviation from the exact value is $1.2\%$).

\item We now introduce a small parameter $\ve$ near the linear term of Eq. (\ref{eq1})
\begin{equation}\label{eq4}
x^5+\ve x=1.
\end{equation}
Presenting the solution of Eq. (\ref{eq4}) in the form    
\begin{equation}\label{eq5}
 x(\ve)=b_0+b_1\ve+b_2\ve^2+\dots,
\end{equation}
we have, after applying the standard procedure of perturbation method\\\\
$
 b_0=1,\quad b_1=-1,\quad b_2=-1\frac{1}{25},\quad a_3=-1\frac{1}{125},\quad b_4=0,\quad b_5=\frac{21}{15625},\quad b_6=\frac{78}{78125}.
$\\\\
And in this case we can construct a general expression for the coefficients with:
\[
 b_n=-\dfrac{\Gamma\left[(4n-1)/5\right]}{5\Gamma\left[(4-n)/5\right]n!},
\]

and determine the radius of convergence of the series (\ref{eq4}): $R=\frac{5}{4^{(4 /5)}}=1,64938\dots$ .  
The value of $x(1)$, taking into account the first six terms of the series (\ref{eq5}), deviates from the exact by $0.07\%$.

\item We introduce the "small parameter" $\delta$ in the exponent
\begin{equation}\label{eq6}
 x^{1+\delta}+x=1,
\end{equation}
and represent $x$ in the form
\begin{equation}\label{eq7}
 x=c_0+c_1\delta+c_2\delta^2+\dots .
\end{equation}
In addition, we use the expansion:
\[
 x^{1+\delta}=x(1+\delta\ln |x|+\dots).
\]

Coefficients of the series (\ref{eq7}) are determined easily:
\[
 c_0=0.5,\quad c_1=0.25 \ln 2,\quad c_2=-0.125\ln 2,\quad \dots .
\]

The radius of convergence is one in this case, and calculate when it should be $\delta=4$.  
Using Pad\'{e} approximants with three terms in the numerator and denominator, if  $\ve=1$, 
we find $x=0.75448$ that only deviates from the exact by 0.05\% result. Calculating $c_i$ for $i=0,1,\dots ,12$  
and constructing Pad\'{e} approximant with six terms in the numerator and denominator, we find $x=0.75487654$ (0.00015\% error).
The method is called "the method of small delta" (see Sect. \ref{Method of Small Delta}).

\item We now assume the exponent as a large parameter. Consider the equation
\begin{equation}\label{eq8}
 x^{n}+x=1.
\end{equation} 
Assuming $n\rightarrow\infty$  (a method of large $\delta$, see Sect. \ref{Method of Large Delta}), we represent the desired solution in the form
\begin{equation}\label{eq9}
 x=\left[\frac{1}{n}(1+x_1+x_2+\dots)\right]^{1/n},
\end{equation}
where  $1>x_1>x_2> \dots .$.

Substituting the ansatz (\ref{eq9}) in Eq. (\ref{eq8}) and taking into account that
\[
 n^{1/n}=1+\dfrac{1}{n}\ln n+\dots,\quad x^{1/n}=1+\dfrac{1}{n}\ln(1+x_1+x_2+\dots)+\dots,
\]
one obtains in order of increasing accuracy of the formula
\begin{eqnarray}
 x&\approx &\left(\frac{\ln n}{n}\right)^{1/n},\label{eq10}\\\nonumber\\
 x&\approx &\left(\frac{\ln n-\ln\ln n}{n}\right)^{1/n},\label{eq11}\\\nonumber\\\nonumber
&\dots&.
\end{eqnarray}
 For  $n = 2$ formula (\ref{eq10}) gives $x = 0.58871$; the error compared to the exact 
solution ($0.5(\sqrt{5}-1)\approx 0.618034$) is 4.7\%. When $n = 5$, we obtain $x=0.79715$ from Eq. (\ref{eq10}) (from the numerical solution one obtains $x=75488$; error 4.4\%). 
For $n = 5$, Eq. (\ref{eq11}) gives $x = 0.74318$ (error 1.5\%). 
Thus, even the first terms of the large $\delta$ asymptotics give excellent results. 

Hence, in this case the method of large delta provides already a good accuracy at low orders of perturbation method.
Approximation (\ref{eq10}), (\ref{eq11}) give an example of nonpower asymptotics.
\end{enumerate}

\CCLsubsection{Homotopy Perturbation Method}

In recent years the so-called homotopy perturbation method was popular \citep{book48} (term ``method of artificial small parameters'' is also used). 
Its essence is as follows. 
In the equations or boundary conditions a parameter $\ve$ is introduced so that for $\ve=0$ one obtains a BVP which admits a simple solution, 
and for $\ve=1$ one obtains the governing BVP. Then the perturbation method to $\ve$ is apllied and in the resulting solution was  adopted. 
Of course, it is not new and was already used by \citet[chapt. I.6]{book33-1} \citet{book51} and \citet{book66-1}. 
The novelty (and, in my opinion, successful) is the title, emphasizing the continuous transition from the initial value $\ve=0$ to the value of $\ve=1$ (homotopy deformation). 
Let us analyse an example of homotopy perturbation parameter method using \citet{book7}. 
A special feature of nonlinear systems with distributed parameters is the possibility of internal resonance between modes. 
That is why in many cases the neglection of higher modes can lead to significant errors. 
The following describes the asymptotic method of solving problems of nonlinear vibrations of systems with distributed parameters, allowing approximately to take into account all modes.
The oscillations of a square membrane on a nonlinear elastic foundation can be written as:
\begin{equation}\label{eq12}
 \dfrac{\partial^2 w}{\partial x^2}+ \dfrac{\partial^2 w}{\partial y^2}- \dfrac{\partial^2 w}{\partial t^2}-cw-\ve w^3=0,
\end{equation}
where $\ve$ is the dimensionless small parameter ($\ve\ll 1$). 

The BCs are as follows:
\begin{equation}\label{eq13}
 \left. w \right|_{x=0,L}= \left. w \right|_{y=0,L}=0.
\end{equation}

       The desired periodic solution must satisfy the periodicity conditions
\begin{equation}\label{eq14}
w(t)=w(t+T),
\end{equation}
where $T=\frac{2\pi}{\omega}$  is the period, and $\Omega$ is the natural frequency of oscillation.
We seek the natural frequencies corresponding to these forms of fundamental vibration frequencies at which the linear case 
($\ve=0$) is realized by one half-wave in each direction $x$ and $y$. We introduce the transformation of time:
\begin{equation}\label{eq15}
\tau=\omega t.
\end{equation}
       The solution is sought in the form of expansions
\begin{equation}\label{eq16}
 w=w_0+\ve w_1+\ve^2 w_2+\dots,
\end{equation}
\begin{equation}\label{eq17}
 \omega=\omega_0+\ve \omega_1+\ve^2 \omega_2+\dots .
\end{equation}
     Substituting ansatzes (\ref{eq16}), (\ref{eq17}) into Eqs. (\ref{eq12}) - (\ref{eq14}) and equating terms of equal powers , we obtain a recurrent sequence of linear BVPs:
\begin{eqnarray}
 \dfrac{\partial^2 w_0}{\partial x^2}+ \dfrac{\partial^2 w_0}{\partial y^2}- \omega_0^2\dfrac{\partial^2 w_0}{\partial \tau^2}-cw_0&=&0,\label{eq18}\\
 \dfrac{\partial^2 w_1}{\partial x^2}+ \dfrac{\partial^2 w_1}{\partial y^2}- \omega_1^2\dfrac{\partial^2 w_1}{\partial \tau^2}-cw_1&=&
2\omega_0\omega_1\dfrac{\partial^2 w_0}{\partial \tau^2}+w_0^3,\label{eq19}\\
&\dots&\nonumber
\end{eqnarray}

The BCs (\ref{eq13}) and periodicity conditions (\ref{eq14}) take the form for $i=1,2,\dots$:
\begin{equation}\label{eq20}
 \left. w_i \right|_{x=0,L}= \left. w_i \right|_{y=0,L}=0,
\end{equation}
\begin{equation}\label{eq21}
w_i(\tau)=w_i(\tau+2\pi).
\end{equation}

        The solution of Eq. (\ref{eq18}) is as follows:
\begin{equation}\label{eq22}
w_{0,0}=\sum\limits_{m=1}^{\infty}\sum\limits_{m=1}^{\infty}
A_{m,n}
\sin\left(\dfrac{\omega_{m,n}}{\omega_0}\tau\right)
\sin\left(\dfrac{\pi m}{L} x\right)
\sin\left(\dfrac{\pi n}{L} y\right),
\end{equation} 
where $\omega_{m,n}=\sqrt{\pi^2\frac{\left(m^2+n^2\right)}{L}+c},\quad m,\;n=1,\;2,\;3,\dots$ and  $A_{1,1}$ is the amplitude of the fundamental tone of vibrations; 
$A_{m,n}$, $m,\;n=1,\;2,\;3\dots$, $(m,n)\neq (1,1)$  is the amplitude of the subsequent modes; 
$\omega_{m,n}$ are the natural frequencies of the linear system, $\omega_0=\omega_{1,1}$.

       The next approximation is a result of solving the BVP (\ref{eq19})-(\ref{eq21}). 
To prevent the appearance of secular terms in the r.h.s. of Eq. (\ref{eq19}) the coefficients of the terms of the form 
\[
 \sin\left(\dfrac{\omega_{m,n}}{\omega_0}\tau\right)
\sin\left(\dfrac{\pi m}{L} x\right)
\sin\left(\dfrac{\pi n}{L} y\right),\quad m,\;n=1,\;2,\;3,\dots
\]
should be equated to zero.

       These conditions lead to an infinite system of nonlinear algebraic equations: 
\begin{equation}\label{eq23}
\dfrac{2A_{m,n}\omega_1}{\beta_2\omega_0}\left(\omega_{m,n}\right)^2=
\sum\limits_{i=1}^{\infty}
\sum\limits_{j=1}^{\infty}
\sum\limits_{k=1}^{\infty}
\sum\limits_{l=1}^{\infty}
\sum\limits_{p=1}^{\infty}
\sum\limits_{s=1}^{\infty}
C_{m,n}^{(ijklps)} 
A_{i,j} A_{k,l} A_{p,s},
\end{equation} 
where $m,n=1,2,3,\dots$.

        Coefficients are found by substituting ansatz (\ref{eq22}) into the r.h.s. of Eqs. (\ref{eq19}) and the relevant simplifications. 
System (\ref{eq23}) can be solved by reduction. However, a sufficiently large number of retention equations beging to show significant computational difficulties. 
In addition, this approach does not take into account the influence of higher oscillation modes. Therefore, further we use the homotopy perturbation method.

       On the right side of each $(m,n)$-th equation of system (\ref{eq23}) we introduce the parameter $\mu$ to those members of  $A_{i,j} A_{k,l} A_{p,s}$, 
for which the following condition is valid: $(i>m)\cup(k>m)\cup(p>m)\cup(j>n)\cup(l>n)\cup(s>n)$. 
Thus, for $\mu=0$ system (\ref{eq23})  takes the "triangular" appearance, and for $\mu=1$ it returns to its original form. Next, we seek a solution in the form of expansions:
\begin{eqnarray}
  \omega_1&=&\omega^{(0)}+\mu\omega^{(1)}+\mu^2\omega^{(2)}+\dots, \label{eq24}\\
 A_{m,n}&=&A_{m,n}^{(0)}+\mu A_{m,n}^{(1)}+\mu^2A_{m,n}^{(2)}+\dots,\label{eq25}
\end{eqnarray}
where $m,n=1,2,3,\dots,\quad (m,n)\neq (1,1)$. 

We put $\mu=1$. 
This approach allows us to keep any number of equations in system (\ref{eq23}). Below we limit ourselves the first two terms
in the expansions (\ref{eq24}), (\ref{eq25}).
We analyze the solutions and note that in this problem the parameter $c$ plays the role of the bifurcation parameter. In general, for $c\neq 0$, $c\sim 1$ , 
the system (\ref{eq23}) admits the following solution:
\[
\begin{array}{cccc}
A_{i,j},\quad i,j=1,2,3,\dots,\quad (i,j)\neq (m,n),\\\\ \omega_1=\dfrac{27}{128}\dfrac{A_{m,n}^2\omega_0}{\omega_{m,n^2}},\quad m,n=1,2,3,\dots.
\end{array}
\]

Amplitude-frequency response is given by:
\[
 \Omega_{m,n}=\omega_{m,n}+0.2109375\dfrac{A_{m,n}^2}{\omega_{m,n}}\ve+\dots .
\]

Of particular interest is the case when the linear component of the restoring force is zero ($c=0$) and the phenomenon of internal resonance between the modes of oscillations occurs. 
Solving system (\ref{eq23}) by the described method, we find\\
$A_{m,n}=0$, $m,n=1,2,3,\dots$, $(m,n)\neq (1,1)$, $(m,n)\neq (2i-1,2i-1)$, $i=1,2,3\dots,\quad A_{3,3}=-4.5662\cdot 10^{-3}A_{1,1}$, $A_{5,5}=2.1139\cdot 10^{-5}A_{1,1}$, $\dots$, 
$\omega_1=0.211048A_{1,1}^2 / \omega_0.$\\

If the oscillations are excited by the mode $(1,1)$ all odd modes $(3,3),\;(5,5)$ etc. are also realized.
However, if the oscillations are excited by one of the higher modes, 
the result of redistribution of energy modes appear at lower orders until the fundamental tone $(1,1)$.

\CCLsubsection{Method of Small Delta}\label{Method of Small Delta}

\citet{book17} proposed an effective method of small $\delta$, which we explain by examples. Let us construct a periodic solution of the following Cauchy problem
\begin{equation}\label{eq26}
 x_{tt}+x^3=0,
\end{equation}
\begin{equation}\label{eq27}
 x(0)=1,\quad x_t(0)=0.
\end{equation}
We introduce a homotopy parameter $\delta$ in Eq. (\ref{eq26}), thus:
\begin{equation}\label{eq28}
 x_{tt}+x^{1+2\delta}=0.
\end{equation}
At the final expression one should put $\delta=1$, but in the process of solving one must suppose $\delta\ll 1$. Then
\begin{equation}\label{eq29}
x^{2\delta}=1+\delta \ln x^2+ 0.5\delta^2 \left(\ln x^2\right)^2+\dots .
\end{equation}
We suppose the solution of Eq. (\ref{eq26}) in the form
\begin{equation}\label{eq30}
x=\sum\limits_{k=0}^\infty \delta^k x_k,
\end{equation}
and carry out the change of independent variable
\begin{equation}\label{eq31}
t=\dfrac{\tau}{\omega},
\end{equation}
where $\omega^2=1+\alpha_1\delta +\alpha_2\delta^2 +\dots$.

The constants $\alpha_i$ ($i = 1,2, \dots$) are determined in the problem solving process.
After substituting ansatzes (\ref{eq29})-(\ref{eq31}) into Eq. (\ref{eq28}) and splitting it with respect to $\delta$, we obtain the following recurrent sequence of Cauchy problems:
\begin{equation}\label{eq32}
 x_{0\tau\tau}+x_0=0,
\end{equation}
\begin{equation}\label{eq33}
x_0(0)=1,\quad x_{0\tau}(0)=0;
\end{equation}
\begin{equation}\label{eq34}
 x_{1\tau\tau}+x_1=-x_0\ln(x^2_0)-\alpha_1 x_{0\tau\tau},
\end{equation}
\begin{equation}\label{eq35}
x_1(0)= x_{1\tau}=0;
\end{equation}
\begin{equation}\label{eq36}
 x_{2\tau\tau}+x_2=-x_1\ln(x^2_1)-2x_1-x_0\left(\ln(x_0^2)\right)^2-\alpha_2 x_{0\tau\tau}-\alpha_1 x_{1\tau\tau};
\end{equation}
\begin{equation}\label{eq37}
x_1(0)= x_{1\tau}=0;
\end{equation}
\[
 \dots .
\]
A Cauchy problem of the zero approximation (\ref{eq32}), (\ref{eq33}) has the solution
.\[
  x_0=\cos\tau.
 \]

In a first approximation, one obtains
\[
  x_{1\tau\tau}+x_1=-\cos\tau\ln(\cos^2\tau)+\alpha_1\cos\tau\equiv L_0.
 \]
The condition of the absence of the secular terms in the solution of this equation can be written as follows:
\[
 \int\limits_0^{\pi /2} L_0\cos t\;{\mbox{d}} t=0,
\]
which is determined by the constant $\alpha_1=1-2\ln 2$.
       The expression for the oscillation period can be written as
\[
 T=2\pi[1+\delta(\ln 2 - 0.5)].
\]

For $\delta = 1$, we have $T = 6.8070$, while the exact value is $T = 7.4164$ (the error of approximate solutions is 8.2\%).
The solution of the Cauchy problems of the next approximation (\ref{eq36}),(\ref{eq37}) gives the value of the period, it practically coincides with the exact one ($T = 7.4111$).

We now consider the wave equation
\begin{equation}\label{eq38}
u_{tt}= u_{xx},
\end{equation}
with nonlinear BCs:
\begin{eqnarray}
u(0,t)&=& 0,\label{eq39}\\
u_x(1,t)+u(1,t)+u^3(1,t)&=& 0.\label{eq40}
\end{eqnarray}
We introduce the parameter $\delta$ in Eq. (\ref{eq40}) as follows:
\begin{equation}\label{eq41}
u_x(1,t)+u(1,t)+u^{1+2\delta}(1,t)= 0.
\end{equation}
In the final expression should we put $\delta = 1$, but in the process of decision we assume $\delta\ll 1$. Then
\begin{equation}\label{eq42}
u^3\equiv u^{1+2\delta}=u\left[1+\delta \ln u^2 + \dfrac{\delta^2}{2}\left(\ln u^2\right)^2+\dots\right].
\end{equation}

         Suppose further the solution of  Eq. (\ref{eq38}) in the form
\begin{equation}\label{eq43}
u=\sum\limits_{k=0}^\infty \delta^k u_k.
\end{equation}

After substituting ansatzes (\ref{eq43}), (\ref{eq41}) into Eqs. (\ref{eq38}), (\ref{eq39}), (\ref{eq41}) and the splitting of the parameter 
$\delta$ we obtain the following recurrent sequence of BVPs:
\begin{equation}\label{eq44}
u_{0\tau\tau}=u_{0xx};
\end{equation}
\begin{equation}\label{eq45}
\mbox{at }x=0,\quad u_0=0;
\end{equation}
\begin{equation}\label{eq46}
\mbox{at }x=1,\quad u_{0x}+2u_0=0;
\end{equation}
\begin{equation}\label{eq47}
u_{0\tau\tau}=u_{0xx}-\sum\limits_{p=0}^1\alpha_{i-p}u_{p\tau\tau};
\end{equation}
\begin{equation}\label{eq48}
\mbox{at }x=0,\quad u_1=0;
\end{equation}
\begin{equation}\label{eq49}
\mbox{at }x=1,\quad u_{1x}+2u_1=-u_0\ln u_0^2;
\end{equation}
\begin{equation}\label{eq50}
u_{0\tau\tau}=u_{0xx}-\sum\limits_{p=0}^2\alpha_{i-p}u_{p\tau\tau};
\end{equation}
\begin{equation}\label{eq51}
\mbox{at }x=0,\quad u_2=0;
\end{equation}
\begin{equation}\label{eq52}
\mbox{at }x=1,\quad u_{2x}+2u_2=-u_1\ln u_0^2-2u_1-0.5u_0 \left(\ln u_0^2\right)^2;
\end{equation}
\[
 \dots ,
\]
         where $\alpha_0=0$.

The solution of the BVP of the zero approximation (\ref{eq44})-(\ref{eq46}) can be written as
\begin{equation}\label{eq53}
u_0=A\sin (\omega_0 x)\sin(\omega_0\tau),
\end{equation}
where the frequency  $\omega_0$  is determined from the transcendental equation
\begin{equation}\label{eq54}
\omega_0=2\tan \omega_0.
\end{equation}
The first few nonzero values of $\omega$ are given in Table \ref{tab1}. 
  \begin{table}[h]
  \begin{center}
  $
  \begin{array}{|c|c|c|c|c|}
  \hline
  \omega_0^{(1)}&\omega_0^{(2)}&\omega_0^{(3)}&\omega_0^{(4)}&\omega_0^{(5)}\\\hline
  2.289&
  5.087&
  8.096&
  11.173&
  14.276
  \\\hline
 \multicolumn{5}{c}{}\\\hline
    \omega_0^{(6)}&\omega_0^{(7)}&\omega_0^{(8)}&\omega_0^{(9)}&\omega_0^{(10)}\\\hline
     17.393&
  20.518&
  23.646&
  26.778&
  29.912\\\hline
  \end{array}
  $
    \end{center}
  \caption{The first few roots of the transcendental Eq. (\ref{eq54}).}
  \label{tab1}
  \end{table}

When $k\rightarrow\infty$  we have the asymptotics: $\omega^{(k)}\rightarrow 0.5 \pi (2k+1)$.

       The BVP problem of the first approximation are as follows:
\begin{equation}\label{eq55}
u_{1xx}-u_{1\tau\tau}
=\alpha_1A\omega_0^2\sin (\omega_0 x)\sin(\omega_0\tau);
\end{equation}
\begin{equation}\label{eq56}
\mbox{at }x=0,\quad u_1=0;
\end{equation}
\begin{equation}\label{eq57}
\mbox{at }x=1,\quad 
u_{1x}+2u_1
=A_1\sin(\omega_0\tau)\left[\ln(A^2\sin^2 \omega_0)+\ln\sin^2(\omega_0\tau)\right];
\end{equation}
where $A_1=-A\sin \omega_0$.

The particular solution of  Eq.  (\ref{eq55}) satisfying the BC  (\ref{eq56}) has the form
\begin{equation}\label{eq58}
 u_1^{(1)}=-0.5\alpha_1 A \omega_0 x \cos(\omega_0 x)\sin(\omega_0\tau).
\end{equation}

We choose the constant $\alpha_1$  in such a way that it compensates the secular term on the r.h.s. of Eq. (\ref{eq57})
\[
 \alpha_1=\dfrac{2R_1}{\omega_0(6+\omega_0^2)}
\]
where  $R_1=\ln (0.25 e A^2 \sin^2 \omega_0)$.

Non-secular harmonics on the r.h.s. of Eq. (\ref{eq57}) give the solution
\begin{equation}\label{eq59}
 u_1^{(2)}=4A_1\sum\limits_{k=2}^\infty T_k\sin(\omega_0 k x) \sin(\omega_0 k \tau)\dfrac{1}{k^2-1},
\end{equation}
where $T_k=1 / [k\omega_0 \cos (k\omega_0)+2\sin (k\omega_0)]$.

The complete solution of the first approximation has the form
\[
 u_1=u_1^{(1)}+u_1^{(2)}.
\]

Assuming $\delta = 1$, we obtain the solution of Eqs. (\ref{eq38})-(\ref{eq40}).

Let us now consider the Schr\"{o}dinger equation \citep{book20}
\begin{equation}\label{eq60}
\Psi_{xx}-x^{2N}\Psi+E\Psi =0,
\end{equation}
\begin{equation}\label{eq61}
\Psi(\pm \infty) =0.
\end{equation}

Here $\Psi$ is the wave function; $E$ is the energy,it plays the role of eigenvalue.
It is shown that the eigenvalue problem (\ref{eq60}),(\ref{eq61}) has a discrete countable spectrum $E_n$, $n=0,1,2,\dots$ \citep{book73}.  
For $N = 2$ the eigenvalue problem (\ref{eq60}), (\ref{eq61}) has an exact solution.  Now let $N$ differ little from 2,
\begin{equation}\label{eq62}
\Psi_{xx}-x^{2+2\delta}\Psi+E\Psi =0.
\end{equation}

Now we use the expansion
\[
x^{2\delta} =1+\delta\ln (x^2)+\dots,
\]			
and we will search for the eigenfunction $\Psi$ and the eigenvalue $E$ in the form of the following series:
\begin{equation}\label{eq63}
\Psi=\Psi_0+\delta\Psi+\delta^2\Psi^2+\dots,
\end{equation}
\begin{equation}\label{eq64}
E=E_0+\delta E+\delta^2 E^2+\dots.
\end{equation}
As a result, after the asymptotic splitting we obtain a recursive sequence of eigenvalue problems
\begin{equation}\label{eq65}
\Psi_{0xx}-x^{2}\Psi_0+E_0\Psi_0 =0,
\end{equation}
\begin{equation}\label{eq66}
\Psi_{1xx}-x^{2}\Psi_1+E_0\Psi_1+E_1\Psi_0 =x^2\Psi_0\ln (x^2),
\end{equation}
\[
 \dots,
\]
\begin{equation}\label{eq67}
|\Psi_i|\rightarrow 0 \quad \mbox{at}\quad |x|\rightarrow\infty,\quad i=1,2,3,\dots.
\end{equation}

The solution of the eigenvalue problem (\ref{eq65}), (\ref{eq67}) has the form
\[
E_0^{(n)}=2n+1,\quad \Psi_0^{(n)}=e^{-x^2/2} H_n(x),\quad n=1,2,3,\dots,
\]
where $H_n(x)$ is the Struve function (\citealp[Sect.12]{book1}).
From the eigenvalue problem  (\ref{eq66}),  (\ref{eq67}) we find
\[
 E_1^{(n)}=
\dfrac{
\int\limits_{-\infty}^{\infty}x^2 e^{-x^2}H_n^2(x)\ln (x^2)\; {\mbox{d}}x
}{\sqrt{\pi}\; 2^n n!}.
\]

For $n = 0$ one obtains $H_0(x) = 1$ and  
\[
 \int\limits_{-\infty}^{\infty}x^2\ln x e^{-x^2}{\mbox{d}}x=\dfrac{\sqrt{\pi}}{8}(2-2\ln 2-C),
\]
   
where  $C = 0.577215\dots$ is the Euler constant. 
Hence
\begin{equation}\label{eq68}
E_0^{(0)}=1+\dfrac{1}{16}(2-2\ln 2 - C)\delta+\dots\;.
\end{equation}
\CCLsubsection{Method of Large Delta}\label{Method of Large Delta}

An alternative method of small delta is the method of large delta, which we demonstrate in an example of a nonlinear equation
\begin{equation}\label{eq69}
x_{tt}+x^n=0,\quad n=3,5,7,\dots\; .
\end{equation}
This equation can be integrated with the functions Cs and Sn, introduced by Liapunov  \citet{book51} (inversions of incomplete beta functions, \citet{book70}). 
Note that much later the same (up to normalization) function have been proposed by Rosenberg, who called them Ateb-functions \citet{book67}. 
However, working with these objects is inconvenient, and therefore the problem of the approximate analytical solution of Eq. (\ref{eq69}) in elementary functions arises.
We construct asymptotics of periodic solutions of  Eq.  (\ref{eq69}) at $n\rightarrow\infty$. Let the initial conditions for Eq.  (\ref{eq69}) be
\begin{equation}\label{eq70}
x(0)=0,\quad \dot x(0)=1.
\end{equation}
        The first integral of the Cauchy problem (\ref{eq69}),(\ref{eq70}) can be written as:
\begin{equation}\label{eq71}
\left(\dfrac{\mbox{d} x}{\mbox{d}t}\right)^2=1-\dfrac{2x^{n+1}}{n+1}.
\end{equation}
      The replacement of the $x=\lambda^{-\lambda /2}$, $\lambda=2 /(n+1)$  , and integration  gives us a solution in the implicit form
\[
 \lambda^{\lambda/2}t=\int\limits_0^{0\le \xi\le 1}\dfrac{{\mbox{d}}\xi}{\sqrt{1-\xi^{2/\lambda}}}.
\]
					
After replacing $\xi=\sin^\lambda \theta$ this implicit solution is transformed into an expression that contains a small parameter in the exponent of the integrand:
\[
\lambda^{\lambda/2}t=\lambda\int\limits_0^{0\le \theta\le \pi/2}\sin^{-1+\lambda}\theta {\mbox{d}}\theta.
\] 
We now consider the integrand  separately:
\[
 \sin^{-1+\lambda}\theta
=\theta^{-1+\lambda}\left(\dfrac{\theta}{\sin\theta}\right)
=\theta^{-1+\lambda}\left[\dfrac{\theta}{\sin\theta}-\lambda\ln\dfrac{\theta}{\sin\theta}+\dots \right].
\]

Expanding this function into a Maclaurin series, one obtains
\[
 \sin^{-1+\lambda}\theta=\theta^{-1+\lambda}+\dfrac{\theta^{-1+\lambda}}{3}+\dots+ O(\lambda).
\]
			
The first term of this expression makes the main contribution of this expression, so in the first approximation we can write
\[
\lambda^{\lambda / 2}t\approx \theta^\lambda, \quad \mbox{i.e.} \quad \theta\approx \lambda^{1 /2}t^{1 /\lambda}.
\]

 In the original variables one obtains

\begin{equation}\label{eq72}
x\approx \lambda^{-\lambda/2}\sin^\lambda\left(\lambda^{1 / 2}t^{1 / \lambda}\right)
\end{equation} 

The solution (\ref{eq72})  should be used on a quarter-period $T$
with
\begin{equation}\label{eq73}
T=4\left(\dfrac{\pi}{2\lambda^{1/2}}\right)^\lambda
\end{equation} 
Let us analyze the solution (\ref{eq72}), (\ref{eq73}). At $n=1$ one obtains the exact values $x=\sin t$, $T=2\pi$ , for $n\rightarrow\infty$   one obtains $T\rightarrow 4$. 
Expanding the r.h.s. of the Eq. (\ref{eq72}) in a series of $t$ and restricting it to the first term, we obtain nonsmooth solution \citep{book66}. 
We estimate the error of the solution (\ref{eq73}). For this we use the expression
\begin{equation}\label{eq74}
\lambda\int\limits_0^{\pi/2}\sin^{-1+\lambda}\;{\mbox{d}}\theta=0.5\lambda B(0.5\lambda,0.5)\equiv A_1,
\end{equation} 
where $B(\dots,\dots)$ is the beta function (\citealp[Sect.6]{book1}).
The approximate value of the integral on the l.h.s. of Eq. (\ref{eq72}) is calculated as follows: $A_2=(\pi /2 )^\lambda$ . 
Numerical comparison of the values $A_1$, $A_2$  and error estimate $\Delta$ is given in Table \ref{tab2}.
  \begin{table}[h]
  \begin{center}
  $
  \begin{array}{|c||c|c|c|c|c|c|c|c|c|c|}
  \hline
  n& 1 & 3 & 5           & \dots &\infty\\\hline\hline
  A_1&\pi^2 &1.30 & 1.20 & \dots& 1\\\hline
  A_2&\pi^2 &1.25 & 1.16 & \dots& 1\\\hline
  \Delta, \%& 0 & 5 & 3 &\dots&\sim 0 \\\hline
  \end{array}
  $
    \end{center}
  \caption{Comparison of exact and approximate solutions.}
  \label{tab2}
  \end{table}

Thus, the first approximation of the asymptotics for $n\rightarrow\infty$ already gives quite acceptable accuracy for practical purposes, even for not very large values of $n$.
Note that the expression (\ref{eq72}) gives an approximation of incomplete beta function (\citealp[Sect.6]{book1}) 
from  $n=1$  (sinus function) to $n=\infty$ (linear function).

\CCLsubsection{Application of Distributions}
Asymptotic methods are based, generally speaking, on the use of Taylor series. 
In this connection the question arises: what to do with the functions of the form $\exp(-\ve^{-1}x)$, 
which cannot be expanded in a Taylor series in $\ve\rightarrow 0$, 
using smooth functions \citep{book65}. The way out lies in the transition to the following distribution \citep{book29}:
\begin{equation}\label{eq75}
 H(x)\exp(-\ve^{-1}x)=\sum\limits_{n=0}^\infty (-1)^n\ve^{n+1}\delta^{(n)}(x),
\end{equation}
where $\delta(x)$ is the Dirac delta function, representing the derivative of the Heavisyde function $H(x)$ in the theory of distribution; 
$\delta^{(n)}(x)$, $n=1,2,\dots$ are the derivatives of the delta function.

 We show how formal the formula (\ref{eq75}) can be obtained. Applying the Laplace transform to function $\exp(-\ve^{-1}x)$, one obtains:
\[
 \int\limits_0^\infty\exp(-\ve^{-1}x)\;{\mbox{d}}x=\dfrac{\ve}{\ve p+1}.
\]	
Expanding the r.h.s. of this equation in a Maclaurin series of $\ve$, and then calculating inverse transform term by term, one obtains expansion (\ref{eq75}). 
Thus, we again use the Taylor series, but now in the dual space.

Here is another interesting feature of the approach using distributions: a singular perturbated problem can be regarded as a regular perturbated one \citep{book29}. 
Suppose, for example,
\[
\ve y'+y=0\quad x>0,\quad y=1\quad \mbox{at}\quad x = 0.
\] 
This is singular perturbated problem: for $\ve=0$ one obtains a smooth solution  $y = 0$, which does not satisfy the given initial condition. 
However, one can seek a solution in the form of a nonsmooth function. Namely, assuming $z(x)=H(x)y(x)$, one obtains 
\begin{equation}\label{eq76}
\ve z'=-z+\ve \delta(x).
\end{equation}
The solution of Eq. (\ref{eq76}) is sought in the form of expansions
\[
z=\sum\limits_{n=0}^\infty z_n \ve^n.
\]
As a result, one obtains
\begin{equation}\label{eq77}
z_0=0,\quad z_1=\delta(x),\quad z_{n+1}=(-1)^n\delta^{(n)},\quad n=1,2,\dots .
\end{equation}
       Note that the expressions (\ref{eq77}) can go to the smooth functions. 
To do this, it is possible to apply the Laplace transform, then  the Pad\'{e} approximants in the dual spase and then to calculate inverse Laplace transform.

      We show other application of the asymptotic method using distribution. Consider the equation of the membrane, reinforced with fibers of the small, but finite width $\ve$.
      The governing PDE is:
\begin{equation}\label{eq78}
[1+2\ve\Phi_0(y)]u_{xx}+u_{yy}=0,
\end{equation}
where $\Phi_0(y)=\sum\limits_{k=-\infty}^\infty [H(y+kb-\ve)+H(y-kb+\ve)]$.

      Let us expand the function $\Phi_0(y)$ in a series of $\ve$. Applying the two-sided Laplace transform \citep{book78}, one obtains:
\[
 \bar \Phi (p,\ve)=\int\limits_{-\infty}^{\infty}e^{-p|y|}\Phi(y,\ve)\;{\mbox{d}}y
\]

    Expanding the function $ \bar \Phi (p,\ve)$  in a series of $\ve$ and performing the inverse Laplace transform, we obtain
\begin{equation}\label{eq79}
\Phi_0(y)=2\ve \Phi (y) +2\ve \sum\limits_{k=1,3,5,\dots}\ve^n\Phi^{(n)}(y),
\end{equation}
where $\Phi(y)=\sum\limits_{k=-\infty}^{\infty}\delta(y-kb)$.

	Now let us consider the solution of the Eq. (\ref{eq78}) in the form
\begin{equation}\label{eq80}
u=u_0+\ve u_1+\ve^2u_2+\dots .
\end{equation}
         Substituting ansatzes (\ref{eq79}), (\ref{eq80}) into Eq. (\ref{eq78}) and splitting the resulting equation with respect to $\ve$, we arrive at the recursive sequence of BVPs
\begin{eqnarray}
\left[1+2\ve\Phi(y)\right]u_{0xx}+u_{0yy}&=&0,\label{eq81}\\
\left[1+2\ve\Phi(y)\right]u_{1xx}+u_{1yy}
&=&-\ve u_{0xx}\Phi_y(y),
\label{eq82}\\
&\dots&\nonumber
\end{eqnarray}


Thus, in the zero approximation, we obtain the problem with one-dimensional fibers  (\ref{eq81}), 
the influence of the width of the filaments is taken into account in the following approximations  (\ref{eq82}).

\CCLsection{Summation of Asymptotic Series}
\CCLsubsection{Analysis of Power Series}

We suppose that one obtains the following series as a result of an asymptotic study:
\begin{equation}\label{eq2-1}
f(\ve)\sim\sum\limits_{n=0}^\infty C_n \ve^n\quad\mbox{for}\quad \ve\rightarrow 0.
\end{equation}
As it is known, the radius of convergence $\ve_0$ of series (\ref{eq2-1}) is determined by the distance to the nearest singularity of the function 
$f(\ve)$ on the complex plane and can be found using the Cauchy-Hadamard formula:
\[
\dfrac{1}{\ve_0}=\overline{\lim_{n \to \infty}}\left| C_n \right|^{1 / n}.
\]
If the nearest singularity lies on the positive real axis, then the coefficients $C_n$ have the usually one and the same algebraic sign, for example,
\[
\dfrac{1}{1-\ve}\sim 1+\ve+\ve^2+\ve^3+\dots .
\]
If the nearest singularity is located on the negative axis, the algebraic signs of the coefficients $C_n$ are usually alternated, for example,
\[
\dfrac{1}{1+\ve}\sim 1-\ve+\ve^2-\ve^3+\dots .
\]
The pattern of signs is usually set pretty quickly. If there are several features of the same radius, 
that could happen to a real function with complex singularities necessarily occurring in complex conjugate pairs, then the rule of alternation of signs may be more complex, such as
\[
\dfrac{1+\ve}{1+\ve^2}\sim 1+\ve-\ve^2-\ve^3+\ve^4+\ve^5-\ve^6-\ve^7 \dots .
\]
Here we have a pattern of signs $++--$. 
To define $\ve_0$ may it is useful to apply the so-called Domb-Sykes plot \citep{book38, book79, book80, book81}. 
Let the function $f$ have one of the nearest singularities at a point $\ve=\pm \ve_0$ with an index of $\alpha$, i.e.
\[
f(\ve)\sim\left\{
 \begin{array}{llll}
  (\ve_0\pm \ve)^\alpha&\mbox{for }&\alpha\neq0,1,2,\dots,\\\\
  (\ve_0\pm \ve)^\alpha\ln(\ve_0\pm \ve)&\mbox{for }&\alpha =0,1,2,\dots ,
 \end{array}
\right.
\]
then we get
\[
 \dfrac{C_n}{C_{n-1}}\sim \mp \dfrac{1}{\ve_0}\left(1-\dfrac{1+\alpha}{n}\right)n
\]

Constructing a graph of $C_n / C_{n-1}$ an the vertical axis and $1 / n$ on the horizontal axis, one obtains the radius of convergence (as the reciprocal of the intercepts on the axis $C_n / C_{n-1}$), 
and then, knowing the slope, the required singularity. Figure 1 shows the numerical results for the function
\begin{equation}\label{eq2-2}
\begin{array}{llll}
f(\ve)&=&\ve(1+\ve)(1+2\ve)^{-1/2}\\\\
&\sim& \ve-\ve^2+\dfrac{3}{2}\ve^3-\dfrac{3}{2}\ve^4+\dfrac{27}{8}\ve^5-\dfrac{51}{8}\ve^6+\dfrac{191}{16}\ve^7-\dfrac{359}{16}\ve^8+\dots,
\end{array}
\end{equation}
starting with $n = 7$ points arranged in a linear relationship.

\begin{figure}[!ht]
	\begin{center}
  	\scalebox{0.55}{\includegraphics*[0pt,140pt][440pt,430pt]{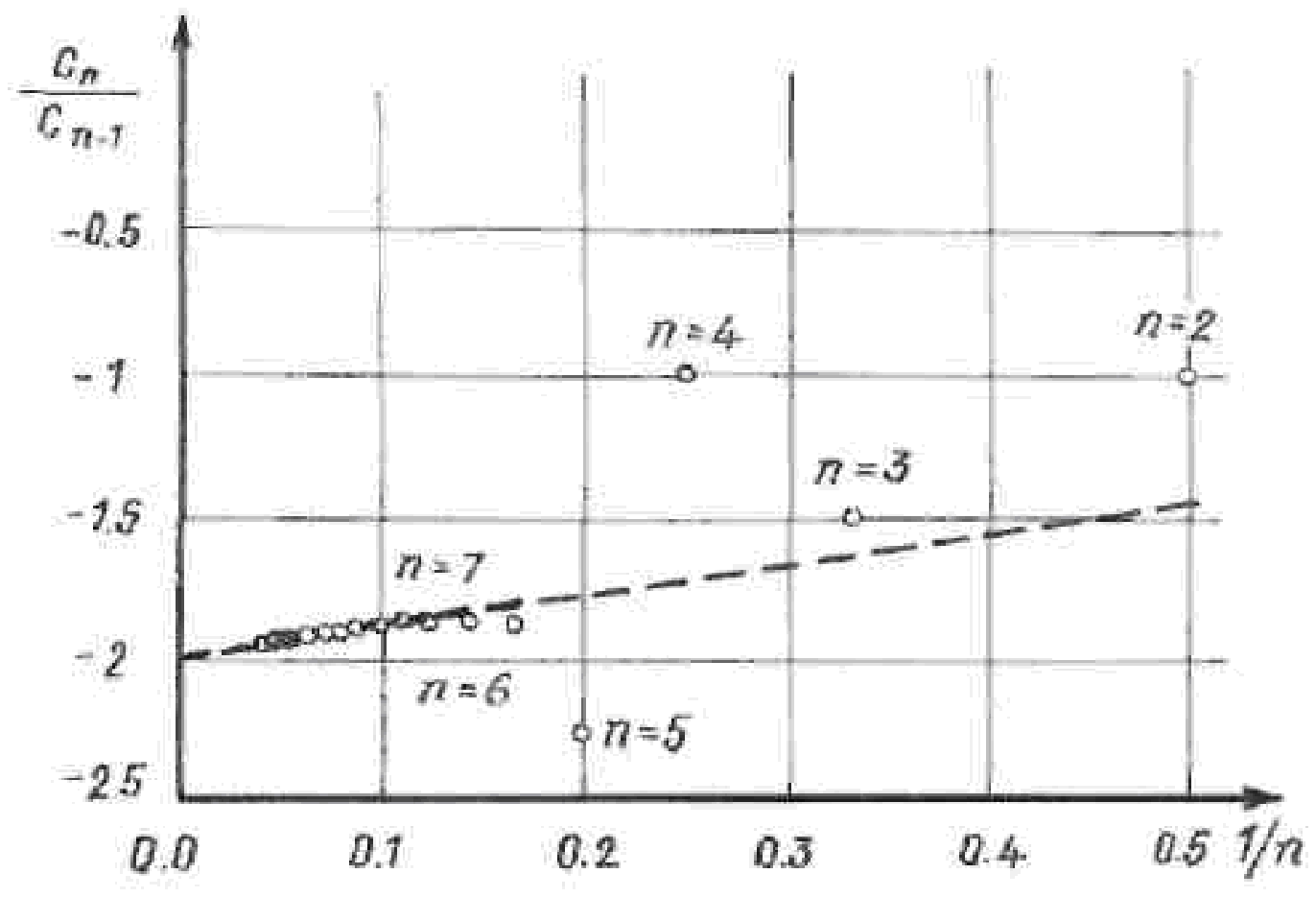}}
	\end{center}
	\vspace*{-0.5cm}
	\caption{The Domb-Sykes plot for $f(\ve) \,=\, \ve (1+\ve )(1+2\ve ) ^{-1/2}$ .}
	\label{pic2.1}
\end{figure}

If $\ve_0$ or $\alpha$ are known from physical considerations, they can be used for the construction of the Domb-Sykes plot.
 If several singularities have the same convergence radius, so that the signs of the coefficients oscillate, 
you can try to construct a dependence on the value  $\left( C_n / C_{n-1} \right)^{1 / 2}$. If the radius of convergence tends to infinity and $C_n / C_{n-1}\sim k/n$, 
then the analyzed function has a factor $\exp(k\ve )$, when $C_n/ C_{n-1}\sim k /n^{1 /p} $ it has a factor $\exp(\ve^p)$. 
If the radius of convergence tends to zero, 
then the analyzed function has an essential singularity and asymptotic expansion diverge. 
If the coefficients behave like  $C_{n-1} / C_{n} \sim 1 / (kn)$ then we can write $C_n\sim C k^n n!$, where $C$ is a constant.

Knowledge of the singular solutions can eliminate them from the perturbation series and thus significantly improve its convergence. 
We describe some techniques for removing singularities. If the singularity lies on the positive real axis, then it often means that the function $f(\ve)$ is multivalued
and that there is a maximum attainable point $\ve=\ve_0$. 
Then the inverse of the original function $\ve=\ve(f)$ can be single valued. For example, consider the function
\begin{equation}\label{eq2-3}
f(\ve)=\arcsin \ve = \ve+\dfrac{1}{6}\ve^3+\dfrac{3}{40}\ve^5+\dfrac{5}{112}\ve^7+\dots,
\end{equation}
the inverse of this function is
\begin{equation}\label{eq2-4}
\ve\sim f-\dfrac{1}{6}f^3+\dfrac{1}120f^5-\dfrac{1}{5040}f^7+\dots\;.
\end{equation}
Numerical results are shown in Figure 2, where the solid line denotes the function $\arcsin\ve$, 
the dotted and dashed line shows the $n$-term expansions (\ref{eq2-3}) and k-terms expansions (\ref{eq2-4}) for different numbers of terms. 
It is evident that the expansion (\ref{eq2-4}) allows a good description of the second branch of the original function.

\begin{figure}[!ht]
	\begin{center}
  	\scalebox{0.55}{\includegraphics*[10pt,100pt][450pt,430pt]{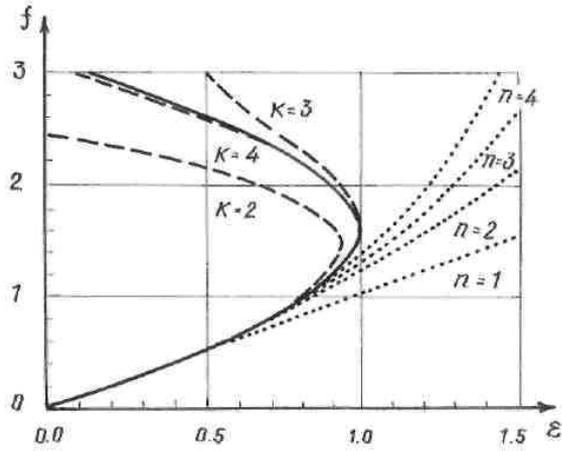}}
	\end{center}
	\vspace*{-0.5cm}
	\caption{Application of the inversion of a power series.}
	\label{pic2.2}
\end{figure}

If
\[
 f\sim A(\ve_0-\ve)^\alpha\quad \mbox{for}\quad \ve\rightarrow\ve_0,\quad 0<\alpha<1,
\]
the transition to the function $f^{1/2}$ removes the singularity.

Consider the following example:
\begin{equation}\label{eq2-5}
f(\ve)=e^{-\ve / 2}\sqrt{1+2\ve}\sim1+\dfrac{1}{2}\ve-\dfrac{7}{8}\ve^2+\dfrac{41}{48}\ve^3-\dfrac{367}{384}\ve^4+\dfrac{4849}{3840}\ve^5+\dots .
\end{equation}

The radius of convergence of this expansion is equal to $1 /2$, while the radius of convergence of functions
\begin{equation}\label{eq2-6}
f^2\sim  1+\ve-\dfrac{3}{2}\ve^2+\dfrac{5}{6}\ve^3-\dfrac{7}{24}\ve^4+\dfrac{3}{40}\ve^5+\dots .
\end{equation}
is infinite.

Numerical results are shown in Figure 3, where the solid line denotes the function $f(\ve)=e^{-\ve / 2}\sqrt{1+2 \ve}$, 
the dotted and dashed lines shows the $n$-term expansions (\ref{eq2-5}) and the square roots of $k$-term expansions (\ref{eq2-6}).
\begin{figure}[!ht]
	\begin{center}
  	\scalebox{0.55}{\includegraphics*[10pt,100pt][450pt,400pt]{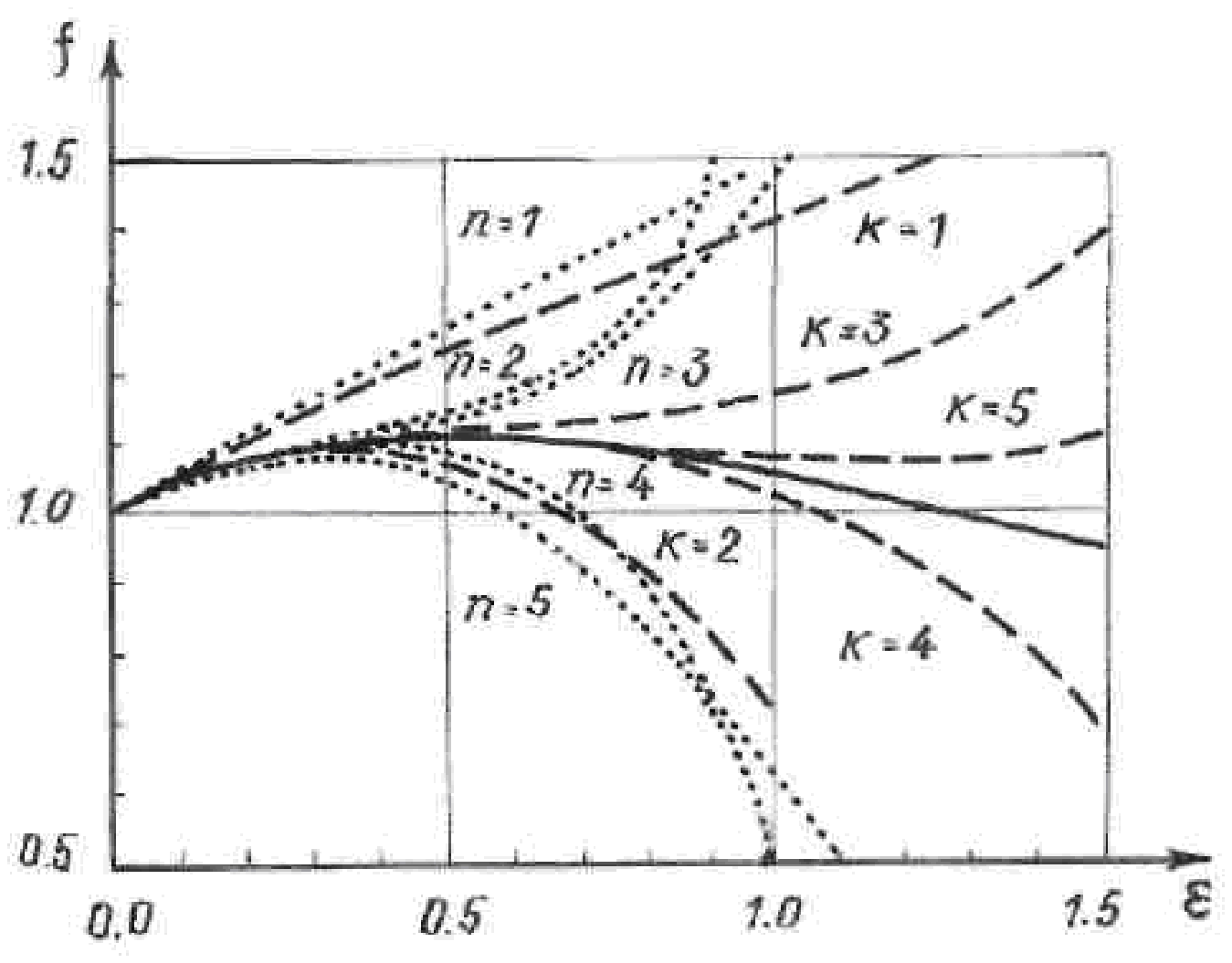}}
	\end{center}
	\vspace*{-0.5cm}
	\caption{An example of taking the root.}
	\label{pic2.3}
\end{figure}

In addition, knowing the singularity, one can construct a new function $f_M(\ve)$  (multiplicative extraction rule)
\[
 f(\ve)=(\ve_0-\ve)^\alpha f_M(\ve),
\]
or $f_A(\ve)$ (additive extraction rule)
\[
 f(\ve)=A(\ve_0-\ve)^\alpha f_A(\ve).
\]

The functions $f_M(\ve )$ and $f_A(\ve )$  should not contain singularities at $\ve_0$.
In many cases, one can effectively use the conformal transformation of the series, a fairly complete catalog of which is given in \citep{book45}. 
In particular, it sometimes turns out to be a successful Euler transformation \citep{book16, book80, book81}, based on the introduction of a new variable
\begin{equation}\label{eq2-7}
\tilde \ve=\frac{\ve}{1-\ve /\ve_0}.
\end{equation}
Recast the function $f$ in terms of $\tilde \ve$ , $f\sim \sum d_n\tilde \ve ^n$, has the singularity pushed out at the point $\tilde \ve=\infty$. 
For example, the function (\ref{eq2-2}) is singular at the $\ve=-1/2$, which can be eliminated with the Euler transformation $\tilde \ve=\ve / (1+2\ve)$. 
The expansion of the function (\ref{eq2-2}) in terms of $\tilde \ve$ is
\begin{equation}\label{eq2-8}
f(\tilde \ve)\sim 1+\dfrac{1}{2}\tilde \ve+\dfrac{1}{8}\tilde \ve^2-\dfrac{31}{48}\tilde \ve^3-\dfrac{895}{384}\tilde \ve^4-\dfrac{22591}{3840}\tilde \ve^5+\dots .
\end{equation}
Some numerical results are shown in Figure 4, where the dotted and dashed line shows the $n$-term expansions (\ref{eq2-2}) and $k$-terms in the expansion (\ref{eq2-8}).

\begin{figure}[!ht]
	\begin{center}
  	\scalebox{0.55}{\includegraphics*[10pt,120pt][450pt,420pt]{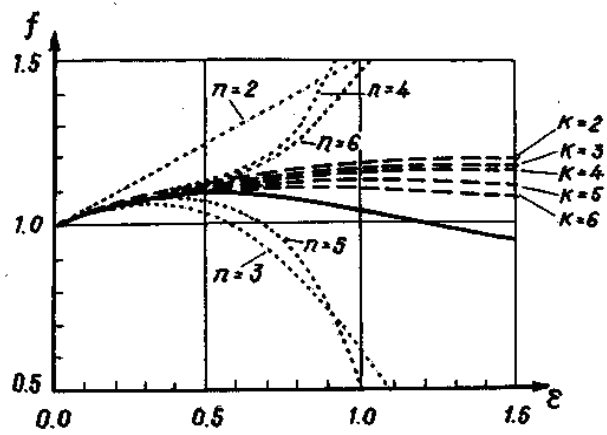}}
	\end{center}
	\vspace*{-0.5cm}
	\caption{Illustration of Euler transformation.}
	\label{pic2.4}
\end{figure}

A natural generalization of Euler transformation is
\[
\tilde \ve=\dfrac{\ve}{(1-\ve/\ve_0)^\alpha},
\]
where  $\alpha$ is the real number.

\CCLsubsection{Pad\'{e} Approximants and Continued Fractions}
\label{Pade Approximants and Continued Fractions}

\textit{''The coefficients of the Taylor series in the aggregate have a lot more information about the values of features than its partial sums. 
It is only necessary to be able to retrieve it, and some of the ways to do this is to construct a Pad\'{e} approximant``} \citep{book84}. 
Pad\'{e} approximants (PA) allow us to implement among the most salient natural transformation of power series in a fractional rational function. 
Let us define a PA following \citet{book13}.
Suppose we have the power series
\begin{equation}\label{eq2-9}
f(\ve)=\sum\limits_{i=1}^\infty c_i\ve^i ,
\end{equation}
Its PA can be written as the expression
\begin{equation}\label{eq2-10}
f_{[n/m]}(\ve)=\dfrac{a_0+a_1\ve+\dots +a_n\ve^n}{1+b_1\ve+\dots +b_m\ve^m},
\end{equation}
whose coefficients are determined from the condition
\begin{equation}\label{eq2-11}
(1+b_1\ve+\dots +b_m\ve^m)(c_0+c_1\ve+c_2\ve^2+ \dots)=a_0+a_1\ve+\dots +a_n\ve^n+O(\ve^{m+n+1}).
\end{equation}
Equating the coefficients of the same powers $\ve$, one obtains a system of linear algebraic equations
\begin{equation}\label{eq2-12}
\begin{array}{llllllllll}
b_mc_{n-m+1}	&+&	b_{m-1}c_{n-m+2}	&+&	c_{n+1}	&=&0;\\
b_mc_{n-m+2}	&+&	b_{m-1}c_{n-m+3}	&+&	c_{n+2}	&=&0;\\
\vdots		&&	\vdots			&&\vdots &=&\vdots\\
b_mc_{n}	&+&	b_{m-1}c_{n+1}		&+&	c_{n+m}	&=&0,\\
\end{array}
\end{equation}
where $c_j=0$ for $j<0$.

The coefficients $a_i$ can now be obtained from the Eqs. (\ref{eq2-11}) by comparing the coefficients of the powers $\ve$:
\begin{equation}\label{eq2-13}
\begin{array}{llllllllll}
a_0&=&c_0;\\
a_1&=&c_1+b_1c_0;\\
\vdots &&\vdots\\
a_n&=&c_n+\sum\limits_{i=1}^pb_ic_{n-i},
\end{array}
\end{equation}
where $p=\mbox{min}(n,m)$.
	Eqs. (\ref{eq2-11}),(\ref{eq2-12}) are called Pad\'{e} equations. In the case where the system (\ref{eq2-12}) is solvable, 
one can obtain the Pad\'{e} coefficients of the numerator and denominator of the PA. Functions $f_{[n/m]}(\ve)$  at different values of $n$ and $m$ form a set, 
which is usually written in the form of a table, called the Pad\'{e} table (Table 3):

  \begin{table}[h]
  \begin{center}
  $
  \begin{array}{|c|c|c|c|c|}
  \hline
  $\backslashbox{$m$}{$n$}$	&0		&1		&2		&\dots\\\hline
  0				&f_{[0/0]}(\ve)	&f_{[1/0]}(\ve)	&f_{[2/0]}(\ve)	&\dots\\\hline
  1				&f_{[0/1]}(\ve)	&f_{[1/1]}(\ve)	&f_{[2/1]}(\ve)	&\dots\\\hline
  2				&f_{[0/2]}(\ve)	&f_{[1/2]}(\ve)	&f_{[2/2]}(\ve)	&\dots\\\hline
  \dots				&\dots		&\dots		&\dots		&\dots\\\hline
  \end{array}
  $
   \end{center}
  \caption{Pad\'{e} table.}
  \label{tab3}
  \end{table}

The terms of the first row of the Pad\'{e} table correspond to the finite sums of the Maclaurin series. In case of $n = m$ one obtains the diagonal PA, 
the most common practice. Note that the Pad\'{e} table can have gaps for those indices $n,\; m$, for which the PA does not exist.
    We note some properties of the PA \citep{book10, book13, book74}:
\begin{enumerate}
\item If the PA at the chosen $m$ and $n$ exists, then it is unique.
\item If the PA sequence converges to some function, the roots of its denominator tend to the poles of the function. 
This allows for a sufficiently large number of terms to determine the pole, and then perform an analytical continuation. 
\item The PA has meromorphic continuation of a given power series functions.
\item The PA on the inverse function is treated the PA function inverse itself. This property is called duality and more exactly formulated as follows. Let
\begin{equation}\label{eq2-14}
q(\ve)=f^{-1}(\ve) \quad\mbox{and}\quad f(0)\neq 0, \quad \mbox{then}\quad q_{[n/m]}(\ve)=f^{-1}_{[n/m]}(\ve),
\end{equation}   
provided that one of these approximations there.
\item Diagonal PA are invariant under fractional linear transformations of the argument. 
Suppose that the function is given by their expansion (\ref{eq2-9}). Consider the linear fractional transformation that preserves the origin $W=\frac{a\ve}{1+b\ve}$, 
and the function $q(W)=f(\ve)$. 
Then $q_{[n/n]}$, provided that one of these approximations exist. In particular, the diagonal PA is invariant concerning Euler transformation (\ref{eq2-7}).
\item Diagonal PA are invariant under fractional linear transformations of functions. Let us analyse a function (\ref{eq2-9}). Let
\[
 q(\ve)=\dfrac{a+bf(\ve)}{c+df(\ve)}.
\]
 If $c+df(\ve)\neq 0$, then
\[
  q_{[n/n]}(\ve)=\dfrac{a+bf_{[n/n]}(\ve)}{c+df_{[n/n]}(\ve)},
\]
provided that there is $f_{[n/n]}(\ve)$. Because of this property infinite values of PA can be considered on a par with the end.
\item The PA can get the upper and lower bounds for $f_{[n/n]}(\ve)$. For the diagonal PA one has the estimate
\begin{equation}\label{eq2-15}
 f_{[n/n-1]}(\ve)      \leq    f_{[n/n]}(\ve)         \le f_{[n/n+1]}(\ve)   .
\end{equation}

Typically, this estimate is valid for the function itself, i.e. $f_{[n/n]}(\ve) $  in Eq. (\ref{eq2-15}) can be replaced by $f(\ve) $ .
\item Diagonal and close to them a sequence of PA often possess the property of autocorrection \citep{book49, book50}. It consists of the following. 
To determine the coefficients of the numerator and denominator of PA have to solve systems of linear algebraic equations. This is an ill-posed procedure,
 so the coefficients of PA can be determined with large errors. However, these errors are in a certain sense of self-consistency, 
the PA can approximate the searching function with a higher accucary.
This is a radical difference between the PA and the Taylor series.
\end{enumerate}

         Autocorrection property is verified for a number of special functions. 
At the same time, even for elliptic functions the so-called Froissart doublets phenomenon arises, 
consisting of closely spaced zeros and poles to each other (but different and obviously irreducible) in the PA. 
This phenomenon is not of numerical nature, but due to the nature of the elliptic function \citep{book74}. Thus, in general, 
having no information about the location of the poles of the PA, but relying solely on the PA (computed exactly as you wish), 
we can not say that you have found a good approximation for the approximated function.

To overcome these defects several methods are suggested, in particular, the smoothing method \citep{book14}. 
Its essence is that instead of the usual-term diagonal PA for complex functions $f_{[n/n]}(\ve)=p_n(\ve)/q_n(\ve)$ the following expression is used
\[
 f_{[n/n]}(\ve)=
 \dfrac{\overline{q_n(\ve)}p_n(\ve)+\overline{q_{n-1}(\ve)}p_{n-1}(\ve)}
{\overline{q_n(\ve)}q_n(\ve)+\overline{q_{n-1}(\ve)}q_{n-1}(\ve)}
\]

      Here $\overline{f}$ denotes complex conjugation of $f$.  

Now consider the question: in what sense can the available mathematical results on the convergence of the PA facilitate the solution of practical problems? 
Gonchar's theorem (Gonchar, 1986) states: if none of the diagonal PA $f_{[n/n]}(\ve) $ has poles in the circle of radius R, 
then the sequence $f_{[n/n]}(\ve) $ is uniformly convergent in the circle to the original function $f(\ve) $. 
Moreover, the absence of poles of the sequence of the $f_{[n/n]}(\ve) $ in a circle of radius $R$ must be original and confirm convergence of the Taylor series in the circle. 
Since for the diagonal PA invariant under fractional linear maps we have $\ve\rightarrow\frac{\ve}{a\ve+b}$, the theorem is true for any open circle containing the point of decomposition, 
and for any area, which is the union of these circles. A significant drawback in practice is the need to check all diagonal PA. 
For example, the following theorem holds \citep{book85}: Suppose the sequence of diagonal Pad\'{e} approximants of the function $w(\ve)$, 
which is holomorphic in the unit disc and has no poles outside this circle. Then this sequence converges uniformly to $w(\ve)$ in the disc $|z|<r_0$, where $0.583 R < r_0 < 0.584 R$.

How can we use these results? Suppose that there are a few terms of the perturbation series and someone wants to estimate its radius of convergence $R$. Consider the interval $[0,\ve_0]$,
 where the truncated perturbation series and  the diagonal PA of the maximal possible order differ by no more than $5\%$ (adopted in the engineering accuracy of the calculations). 
If none of the previous diagonal PA does not have  poles in a circle of radius $\ve_0$, then it is a high level of confidence to assert that $R\geq\ve_0$.

The procedure of constructing the PA is much less labor intensive than the construction of higher approximations of the perturbation theory. 
The PA is not limited to power series, but to the series of orthogonal polynomials. PA is locally the best rational approximation of a given power series. 
They are constructed directly on its odds and allow the efficient analytic continuation of the series outside its circle of convergence, 
and their poles in a certain sense localize the singular points (including the poles and their multiplicities) 
of the continuation function at the corresponding region of convergence and on its boundary. 
This PA fundamentally different from rational approximations to (fully or partially) fixed poles, including those from the polynomial approximation, 
in which case all the poles are fixed in one, infinity, the point. 
Currently, the PA method is one of the most promising non-linear methods of summation of power series and the localization of its singular points. 
Including the reason why the theory of the PA turned into a completely independent section of approximation theory, 
and these approximations have found a variety of applications both directly in the theory of rational approximations, and in perturbation theory.
Thus, the main advantages of PA compared with the Taylor series as follows:
\begin{enumerate}
 \item Typically, the rate of convergence of rational approximations greatly exceeds the rate of convergence of polynomial approximation. For example, 
the function $e^\ve$  in the circle of convergence approximated by rational polynomials $P_n(\ve)/Q_n(\ve)$ in $4^n$  times better than an algebraic polynomial of degree $2n$.
 More tangible it is property for functions of limited smoothness. Thus, the function $|\ve|$ on the interval $[-1,1]$ can not be approximated by algebraic polynomials,
 so that the order of approximation was better than $1/n$, where $n$ is the degree of polynomial. PA gives the rate of convergence $\sim\exp\left(-\sqrt{2n}\right)$.
\item Typically, the radius of convergence of rational approximation is a large compared with power series. 
Thus, for the function $\arctan (x)$ Taylor polynomials converge only if $| \ve | \leq 1$, and AP - 
everywhere in $C \setminus ((- i\infty, - i] \cup [i, i\infty))$.
\item PA can establish the position of singularities of the function.
\end{enumerate}
Similarly, the PA method is a method of continued fractions \citep{book40}. 
There are several types of continued fractions. The regular $C$-fraction has the form of an infinite sequence, in which $N$-th term can be written as follows
\begin{equation}\label{eq2-16}
 f_N(\ve)=a+
\dfrac{c_0}
{1+\dfrac{c_1\ve}{
1+\dfrac{c_2\ve}{
\begin{array}{lll}
 1+\\
\vdots\\
\dfrac{c_{N+1}\ve}{1+c_N\ve}
\end{array}
}}}.
\end{equation}

The coefficients $c_i$  are obtained after the decomposition of expression (\ref{eq2-16}) in a Maclaurin series and then equating the coefficients of equal powers of $\ve$. 
When $a = 0$ one obtains the fraction of Stieltjes or $S$-fraction. For the function of Stieltjes
\[
 S(\ve)=\int\limits_0^\infty \dfrac{\exp(-t)}{1+\ve t}\;{\mbox{d}}t,
\]
the coefficients of expansion (\ref{eq2-16}) have the form: $a=0$, $c_0=1$, $c_{2n-1}=c_{2n}=n$, $n\geq 1$.

Description of the so-called J-, T-, P-, R-, g-fractions, algorithms for their construction and the range of applicability are described in detail in \citet{book40}.

Continued fractions are a special case of continuous functional approximation \citep{book18}. 
This is the sequence in which the $(n +1)$-th term $c_n(\ve)$ has the form $n$-th iteration of a function $F(\ve)$. 
For the Taylor series one obtains $F(\ve)=1+\ve$, for the continuous fraction $F(\ve)=\frac{1}{1+\ve}$. If $F(\ve)=\exp(\ve)$ one obtains a continuous exponential approximation
\[
 c_n(\ve)=a_0\exp\left\{a_1\ve\exp\left[a_2\ve\dots \exp\left(a_n\ve\right)\right]\right\},
\]
for   $F(\ve)=\sqrt{1+\ve}$
\[
 c_n(\ve)=a_0\sqrt{1+a_1\ve\sqrt{1+a_2\ve\sqrt{1+\dots a_{n-1}\ve\sqrt{1+a_n\ve}}}},
\]
for    $F(\ve)=\ln {1+\ve}$
\[
 c_n(\ve)=a_0\ln\left\{a_1\ve\ln\left[a_2\ve\dots \ln\left(a_n\ve\right)\right]\right\}.
\]

In some cases, such approximations can converge significantly faster than power series.

As an example, we note the solution of the transcendental equation
\[
 x=\ve \ln x
\]                         		     
for large values (\citealp[\S 3.4.9]{book12})) $\lambda$:
\[
 x_0=\ve\ln \ve;\quad x_1=\ve\ln(\ve\ln\ve);\quad x_2=\ve\ln[\ve\ln(\ve\ln\ve)];\dots .
\]

\CCLsection{Some Applications of Pad\'{e} Approximants}
\CCLsubsection{Accelerating Convergence of Iterative Processes}

The efficiency of PA or other methods of summation depends largely on the availability of higher approximations of the asymptotic process. 
Sometimes they can be obtained by using computer algorithms \citep{book60}, 
but in general it remains an open question. Iterative methods are significantly easier to implement. 
As a result of an iterative procedure a sequence of $S_n$ is obtained. 
Suppose that it converges and has the limit value. We introduce the parameter $a$ obtained by the ratio 
\[
 a=\lim_{n \to \infty}\dfrac{S_{n+1}-S_n}{S_n-S}.
\]

It's called superlinear convergence, if $a=0$, a linear for $a<1$ and logarithmic at $a=1$. 
The biggest issues are, of course, logarithmically convergent sequences - alas, widespread practice. 
Very often linearly convergent sequences are also a problem. Therefore, it is often necessary to improve the convergence. 
One method of improving the convergence is to move to a new sequence $T_n$ with the aid of a transformation so that 
\[
 \lim_{n \to \infty}\dfrac{T_{n}-S_n}{S_n-S}=0.
\]
In such cases we say that the sequence $T_n$ converges faster than sequence $S_n$. 
There are linear and nonlinear methods to improve convergence. Linear methods are described by formulas
\[
 T_n=\sum\limits_0^\infty a_{ni}S_i,\; n=0,1,2,\dots
\]
where the coefficients $a_{ni}$ do not depend on the terms of the sequence $S_n$ constant.

Since linear methods improve the convergence of a restricted class of sequences, 
currently nonlinear methods the most popular ones. Among them thr Aitken method \citep{book13} stands out for its easiness, which described by the formula
\begin{equation}\label{eq3-1}
 T_n=S_n-\dfrac{\left(S_{n+1}-S_n\right)\left(S_{n}-S_{n-1}\right)}{S_{n+1}-2S_n+S_{n-1}},\quad n=0,1,2,\dots\;.
\end{equation}

The Aitken method accelerates the convergence of all linear and many of logarithmically convergent sequences. It is very easy to calculate, and in some cases it can be applied iteratively.
A natural generalization of the Aitken transformation is the Shanks transformation \citep{book71}
\begin{equation}\label{eq3-2}
 T_p^{sh}=\dfrac{D_{kp}^{(1)}}{D_{kp}^{(1)}},
\end{equation}

where
\[
\begin{array}{lllll}
 D_{kp}^{(1)}=\left|
\begin{array}{lllllll}
S_{p-k}	&S_{p-k+1}	&\dots &S_{p}\\
\Delta S_{p-k}	&\Delta S_{p-k+1}&\dots &\Delta S_{p}\\
\dots &\dots &\dots &\dots &\\
\Delta S_{p-1}	&\Delta S_{p}&\dots &\Delta S_{p+k-1}\\
\end{array}
\right|,\\\\
D_{kp}^{(2)}=\left|
\begin{array}{lllllll}
1	&1	&\dots &1\\
\Delta S_{p-k}	&\Delta S_{p-k+1}&\dots &\Delta S_{p}\\
\dots &\dots &\dots &\dots &\\
\Delta S_{p-1}	&\Delta S_{p}&\dots &\Delta S_{p+k-1}\\
\end{array}
\right|,\\\\
\Delta S_k=S_{k+1}-S_k
\end{array}
\]

Eq. (\ref{eq3-2}) is called the Shanks transformation of the order $k$ of the sequences $S_k$ to the sequence $T_k$. For $k=1$ one obtains the Aitken transform (\ref{eq3-1}).
Shanks method requires the calculation of determinants, which is not always easy. One can use also Wynn algorithm, which described by the formulas:
\begin{equation}\label{eq3-3}
 T_{-1}^{(n)}=0,\quad T_0^{(n)}=S_n,\quad T_{k+1}^{(n)}=T_{k+1}^{(n+1)}+\dfrac{1}{T_{k}^{(n+1)}-T_{k}^{(n)}}.
\end{equation}
The Wynn algorithm is related to the transformation of Shanks (\ref{eq3-2}) in the following way:
\[
 T_{2k}^{(n)}=T_{k}^{(sh)}(S_n),\quad T_{2k+1}^{(n)}=\dfrac{1}{T_{k}^{(sh)}}(\Delta S_n).
\]

The Wynn algorithm is a quadratic convergent method for solving systems of nonlinear equations \citep{book13}.
There are many other techniques for accelerating of sequences' convergence. One can use them consistently, for example, 
to convert the original sequence into a linearly convergent one, and then apply the method of Aitken. 
One can also use different methods to improve convergence, at each stage by comparing the obtained results \citep{book21}.
All the described methods have a close relationship with the PA. The Aitken method corresponds to the PA $[n/1]$ , the Shanks method  to the PA $[p/k]$, 
for the method of Wynn one obtains $T_{2k}^{(n)}=[n+k/k]$.

\CCLsubsection{Removing Singularities and Reducing the Gibbs Effect}

Consider the problem of uniform plane flow of an incompressible inviscid fluid pasting a thin elliptic airfoil $(|x|\leq 1,\; |y|\leq\ve,\;\ve\ll 1).$ 
The expression for the relative velocity q* of the flow is such \citep[\S 4.4]{book81}:
\begin{equation}\label{eq3-4}
q^* =\dfrac{q}{V}=\dfrac{(1+\ve)\sqrt{1-x^2}}{\sqrt{1-x^2(1+\ve^2)}},
\end{equation}
where $V$ is the free-stream speed.

         The splitting of the r.h.s. of Eq. (\ref{eq3-4}) in a series of $\ve$ can be expressed as:
\begin{equation}\label{eq3-5}
q^*(x,\ve) =1+\ve-\dfrac{1}{2}\ve^2\dfrac{x^2}{1-x^2}-\dfrac{1}{2}\ve^3\dfrac{x^2}{1-x^2}+\dots\; .
\end{equation}

This expression diverges at $x = 1$. No wonder it is not: 
the expansion (\ref{eq3-5}) is obtained as a result of the limiting process $\ds\lim_{\ve \to 0} q(x,\ve),\;x> 1$,
and to get the value of  $q(1,\ve)$, it is necessary to perform the limit as $\ds \lim_{x \to 1} q(x,\ve)$ for $\ve> 0$. 
Divergence of series (\ref{eq3-5}) when $x=1$ indicates that the limit processes cannot be interchanged. 
Now let us apply PA to the r.h.s. of Eq. (\ref{eq3-5})  and then pass to the limit $x\rightarrow 1$. 
After trying various options, we conclude that the best result is given by the PA
\begin{equation}\label{eq3-6}
q^*(x,\ve) =\dfrac{(1-x^2)(1+\ve)}{1-x^2}.
\end{equation}

Numerical results for $\ve = 0.5$ are shown in Figure 5, where the dashed line denotes the solution (\ref{eq3-5}), curves 1 and 2 the exact solution (\ref{eq3-6}) and the PA (\ref{eq3-6}). 
It is seen that the use of PA significantly improves the accuracy of the approximate solution.

\begin{figure}[!ht]
	\begin{center}
  	\epsfig{figure=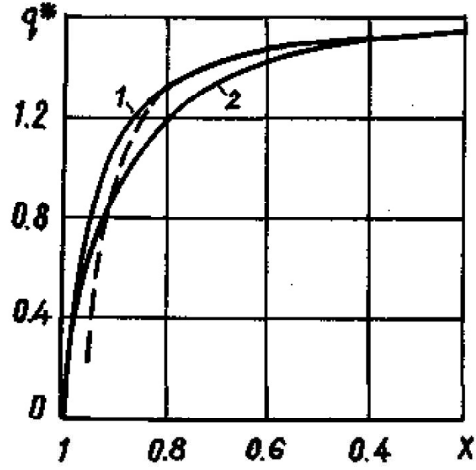,scale=0.25}
	\end{center}
	\vspace*{-1.5cm}
	\caption{Removing singularities by the PA.}
	\label{Fig5}
\end{figure}

PA can be also successfully applied for the suppression of the Gibbs phenomenon \citep{book15, book21, book28, book64} . 
Consider, for example, the function $\mbox{sign }x$:
\[
 \mbox{sign }x =
\left\{
\begin{array}{rrrrrrr}
 -1,&-\pi&<&x&<&0,\\\\
  1,&0&<&x&<&\pi .
\end{array}
\right.
\]

Its Fourier series expansion has the form
\begin{equation}\label{eq3-7}
 \mbox{sign }x =\dfrac{4}{\pi}\sum\limits_{j=0}^\infty \dfrac{\sin(2j+1)x}{2j+1}
\end{equation}

Direct summation of series (\ref{eq3-7}) leads to the Gibbs effect in the neighborhood of $x=0$, 
while the defect of convergence reaches $18\%$, i.e. instead of 1 one obtains the value of $1.1789797\dots $.
Diagonal PA for series (\ref{eq3-7}) can be written as follows:
\begin{equation}\label{eq3-8}
 \mbox{sign }x_{[N/N]} = \dfrac{\sum\limits_{j=0}^{N-1/2}q_{2j+1}\sin\left((2j+1)x\right)}
{1+\sum\limits_{j=0}^{N/2}s_{2j}\cos(2jx)}
\end{equation}
where 
\[
 \begin{array}{lllll}
  q_{2j+1}&=&\dfrac{4}{\pi}(2j+1)\left[\dfrac{1}{(2j+1)^2}+\sum\limits_{i=1}^{[N/2]}\dfrac{s_{2i}}{(2j+1)^2-(2i)^2}\right],\\\\
 s_{2i}&=&2(-1)^i\dfrac{(N!)^4 (2N+2i)!(2N-2i)!}{(N-1)!(N+1)!(N-2i)!(N+2i)![(2N)!]^2}.
 \end{array}
\]

Numerical studies show that the Gibbs effect for PA (\ref{eq3-8}) does not exceed $2\%$ \citep{book64}.

\CCLsubsection{Localized Solutions}
We consider the stationary Schr\"{o}dinger equation
\begin{equation}\label{eq3-9}
\nabla^2(x,y)-u(x,y)+u^3(x,y)=0.
\end{equation}
We seek the real, localized axisymmetric solutions of the Eq.  (\ref{eq3-9}). 
In polar coordinates $(\xi,\theta)$  we construct a solution $\varphi(\xi)$ which does not depend on $\theta$. As a result, we obtain the BVP
\begin{eqnarray}
\varphi'' (\xi )+\dfrac{1}{\xi}\varphi'(\xi)-\varphi(\xi)+\varphi^3(\xi)&=&0,\label{eq3-10}\\
\varphi(x)&=&0,\label{eq3-11}\\
	\lim_{\xi \to \infty}\varphi(\xi)&=&0.	\label{eq3-12}
\end{eqnarray}

        BVP (\ref{eq3-10})-(\ref{eq3-12}) can be regarded as an eigenvalue problem, and the role of an eigenvalue is an unknown quantity $A=\varphi(0)$. 
This problem plays an important role in nonlinear optics, quantum field theory, the theory of magnetic media. 
As shown in \citep[p.12-16]{book61}, BVP (\ref{eq3-10})-(\ref{eq3-12}) has a countable set of "eigenvalues" $A_n$, 
the solution $\varphi(\xi,A_n)$ has exactly $n$ zeros, and the solution $\varphi(\xi,A_0)$ has no zeros and decreases monotonically on $\xi$. 
That is the last solution, which is most interesting from the standpoint of physical applications, and we will concentrate on obtaining it. 

The problem of computing the decaying solutions of BVP (\ref{eq3-10})-(\ref{eq3-12}) is identical to the problem of computing 
homoclinic orbits in the 3D phase space for the nonlinear oscillator, or equivalently, for computing the initial conditions for these orbits.  

Since the sought solutions are expected to be analytical functions of $\xi$, they can be expressed in Maclaurin series about  $\xi=0$:
\begin{equation}\label{eq3-13}
\varphi(\xi)=A_0+\sum\limits_{j=1}^\infty C_{2j}\xi^{2j}.
\end{equation}

Substituting ansatz (\ref{eq3-13}) into Eq. (\ref{eq3-10}), producing a splitting of the powers of the $\xi$ and solving the relevant equations, one obtains \citep{book27}:\\\\
$
C_2=0.25A_0(1-A_0^2);\quad C_4=0.25 C_2 \tilde C;\quad C_6=\dfrac{C_4}{6}-\dfrac{3A_0^2 \tilde C^2}{16};\quad C_8=\dfrac{1}{64}(\tilde C C_6 - 6A_0 C_2C_4-C_2^3);\\\\
C_{10}=-0.01(\tilde C C_8 + 6A_0 C_2 C_6 + 3 A_0 C_4^2 + 3C_2^2 C_4);\\\\
C_{12}=-\dfrac{1}{144}(-\tilde C C_{10}+6A_0 C_2 C_8 + 6 A_0 C_4 C_6+ 3 C_2c_4^2),
$\\\\
where $\tilde C = 1-3A_0^2$.

Then we construct PA for the truncated series (\ref{eq3-13}) 
\begin{equation}\label{eq3-14}
\varphi(\xi)=\dfrac{A_0+\sum\limits_{j=0}^Na_{2j}\xi^{2j}}{1+\sum\limits_{k=1}^N b_{2j}\xi^{2k}}
\end{equation}

     All coefficients in Eq. (\ref{eq3-14}) can be parameterized in terms of $A_0$, $a_{2j}=a_{2j}(A_0)$, $b_{2j}=b_{2j}(A_0)$ and the
PA becomes a one-parameter family of analytical approximations of the searching solution. 
Then we compute the value of $A_0$  for which PA (\ref{eq3-14}) decays to zero as $\xi$ tends to infinity. It gives us condition   
\begin{equation}\label{eq3-15}
 a_{2j}(A_0)=0,\quad  b_{2j}(A_0)\neq 0,
\end{equation}

 One can compute the PA (\ref{eq3-14}) , then imposed the condition (\ref{eq3-15})  and obtained the following convergent values of $A_0$ for varying orders $2N$: 

  \[
  \begin{array}{|c|c|c|c|c|}
  \hline
  N			&1		&2		&3		&4\\\hline
  A_0				&\pm\sqrt{3}	&\pm 2.20701	&\pm 2.21121	&\pm 2.21200\\\hline
  \end{array}
  \]
%
%
%

The numerical solution gives $A_0\approx \pm 2.206208$, the difference between numerical and analytical solutions for $N = 4$ is $0.26\%$. 

\CCLsubsection{Hermite-Pad\'{e} Approximations and Bifurcation Problem}

PA can successfully work with functions having poles. 
However, it often becomes necessary to explore functions with branch points, and construct all their branches. 
In that case, one can use Hermite-Pad\'{e} approximations \citep{book26}. Suppose it comes to a function  with the expansion
\begin{equation}\label{eq3-16}
 f(\ve)=\sum\limits_{n=1}^\infty u_n\ve^n,
\end{equation}
and we managed to find the first few coefficients of this series
\[
 f_N(\ve)=\sum\limits_{n=1}^N u_n\ve^n.
\]		
If it is known that this function has a branch point, we can try to transform the original series (\ref{eq3-16}) in an implicit function
\[
 F(\ve,f)=0,
\]				
and determine all required branches of it.

For this purpose we construct a polynomial $F_p(\ve,f)$  of degree $p\geq 2$
\[
 F_p(\ve,f)=\sum\limits_{m=1}^p \sum\limits_{k=0}^m C_{m-k,k}\ve^{m-k}f^k.
\]
					
 It was assumed $C_{0.1}=1$, and the remaining coefficients must be determined from the condition
\begin{equation}\label{eq3-17}
F_p(\ve, f_N(\ve))=O(\ve^{N+1})\quad \mbox{at}\quad \ve\rightarrow 0.
\end{equation}

Polynomial $F_p$ contains $0.5(p^2+3p-2)$ unknowns, the condition (\ref{eq3-17}) yields $N$ linear algebraic equations, therefore, should $N=0.5(p^2+3p-2)$. 
Once the polynomial $F_p$ is found, one can easily find $p$ branches of the solution from the equation
\[
 F_p=0.
\]

For the analysis of bifurcations of these solutions one can use Newton's polygon \citep{book77}. 
If a priori information about the searching function is known, it can be taken into account for constructing the polynomial $F_p$. 

\CCLsubsection{Estimates of Effective Characteristics of Composite Materials}

We consider a macroscopically isotropic 2D composite material consisting of a matrix with inclusions. 
The aim is to determine the effective conductivity $q$ from the known matrix ($q_1$) and inclusions ($q_2$) conductivities and the volume fraction ($\varphi$). 
As shown in \citep{book76}, if we take  $\ve=\frac{q_2}{q_1}-1$ as a small parameter value, the required effective conductivity can be written as follows
\begin{equation}\label{eq3-18}
\dfrac{q}{q_1}=1+\varphi\ve-0.5\varphi(1-\varphi)\ve^2+O(\ve^3).
\end{equation}

Using the first two terms of expansion (\ref{eq3-18}), one obtains
\[
 \dfrac{1}{1+\varphi\ve}\leq \dfrac{q}{q_1}\leq 1+\varphi\ve,
\]
hence the bounds of Wiener 
\[
 \left(\dfrac{1-\varphi}{q_1}+\dfrac{\varphi}{q_2}\right)^{-1}\leq q \leq (1-\varphi) q_1+\varphi q_2.
\]

\CCLsubsection{Continualization}

      We study a chain of $n+2$ material points with the same masses $m$, 
located in equilibrium states in the points of the axis $x$ with coordinates $jh ( j = 0, 1,\dots , n, n+1)$ and suspended by elastic couplings of stiffness $c$ (Figure 6) \citep{book6}.

\begin{figure}[!ht]
  	\epsfig{figure=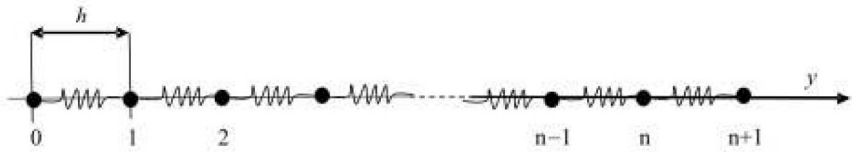,scale=0.23}
	\vspace*{-4.5cm}
	\caption{A chain of elastically coupled masses.}
	\label{Fig6}
\end{figure}

       Owing to the Hooke's law the elastic force acting on the j-th mass is as follows 
\[
 \sigma_j(t)=c[y_{j+1}(t)-y_j(t)]-c[y_{j}(t)-y_{j-1}(t)]=c[y_{j-1}(t)-2y_j(t)+y_{j+1}(t)],
\]
where $j=1,2,\dots, n$ and $y_j(t)$ is the displacement of the $j$-th material point from its static equilibrium position. 

       Applying the second Newton's law one obtains the following system of ODEs governing chain dynamics 
\begin{equation}\label{eq3-28}
m\sigma_{jtt}(t)=c (\sigma_{j+1}-2\sigma_j+\sigma_{j-1}),\quad j=1,2,\dots, n.
\end{equation}
       Let us suppose the following BCs
\begin{equation}\label{eq3-29}
\sigma_0(t)=\sigma_{n+1}(t)=0.
\end{equation}
       For large values of $n$ usually continuum approximation of discrete problem is applied. In our case it takes the form of
\begin{eqnarray}
 m\sigma_{tt} (x,t)&=&ch^2\sigma_{xx}(x,t),\label{eq3-30}\\
 \sigma (0,t)&=&\sigma(\ell,t)=0.\label{eq3-31}
\end{eqnarray}
       Formally, one can rewrite Eq. (\ref{eq3-28}) as pseudo-differential equation: 
\begin{equation}\label{eq3-32}
m\dfrac{\partial^2 \sigma}{\partial t^2}+4c\sin^2\left(-\dfrac{ih}{2}\dfrac{\partial}{\partial x}\right)\sigma=0.
\end{equation}
 Pseudo-differential operator can be split into the Maclaurin series  as follows 
\begin{equation}\label{eq3-33}
\begin{array}{llll}
\sin^2\left(-\dfrac{ih}{2}\dfrac{\partial}{\partial x}\right)
&=&
-\dfrac{1}{2}\sum\limits_{k=1}\dfrac{h^{2k}}{(2k)!}\dfrac{\partial^{2k}}{\partial x^{2k}}\\\\
&=&-\dfrac{h^2}{4}\dfrac{\partial^2}{\partial x^2}
\left(1+\dfrac{h^2}{12}\dfrac{\partial^2}{\partial x^2}+\dfrac{h^4}{360}\dfrac{\partial^4}{\partial x^4}+\dfrac{h^6}{10080}\dfrac{\partial^6}{\partial x^6}\right).
\end{array}
\end{equation}

      With only keeping the first term in the last line of Eq. (\ref{eq3-33}), 
one obtains a continuous approximation (\ref{eq3-30}). Keeping the first three terms in Eq. (\ref{eq3-33}), 
the following model is obtained 
\begin{equation}\label{eq3-34}
m\dfrac{\partial^2 \sigma}{\partial t^2}=
ch^2\left(\dfrac{\partial^2}{\partial x^2}+\dfrac{h^2}{12}\dfrac{\partial^4}{\partial x^4}+\dfrac{h^4}{360}\dfrac{\partial^6}{\partial x^6}\right)
\end{equation}
     In the case of periodic BCs for a discrete chain one obtains the following BCs for Eq. (\ref{eq3-34}):
\begin{equation}\label{eq3-35}
\sigma = \sigma_{xx}=\sigma_{xxxx}=0\quad \mbox{for}\quad x=0,\ell.
\end{equation}
BVP (\ref{eq3-34}), (\ref{eq3-35}) is of the 6th order in spatial variable. Using PA we can obtain a modified continuous approximation of the 2nd order. 
If only two terms are left in in the last line of Eq. (\ref{eq3-33}), then the PA can be cast into the following form:
\[
 \dfrac{\partial^2}{\partial x^2}+\dfrac{h^2}{12}\dfrac{\partial^4}{\partial x^4}\approx 
\dfrac{\dfrac{\partial^2}{\partial x^2}}{1-\dfrac{h^2}{12}\dfrac{\partial^2}{\partial x^2}}.
\]

For justification of this procedure Fourier or Laplace transforms can be used.

The corresponding so-called quasicontinuum model reads 
\begin{equation}\label{eq3-36}
m\left(1-\dfrac{h^2}{12}\dfrac{\partial^2}{\partial x^2}\right)\sigma_{tt}-ch^2\sigma_{xx}=0
\end{equation}

The BCs for Eq. (\ref{eq3-36}) have the form (\ref{eq3-31}).

\CCLsubsection{Rational interpolation}

     Here we follow \citep{book30}. 
     A simple way to approximate a function is to choose a sequence of points
\[
 a=x_0 < x_1 < x_2 \dots < x_n=b,
\]
and to construct the interpolating polynomial $p_n(x)$ 
\[
 p_n(x_i)=f(x_i),\quad i=0,1,2,\dots,n.
\]

     However, as is well-known $p_n(x)$ may not be a good approximation to $f$, and for large $n\gg 1$ it can exhibit wild oscillations. 
If we are free to choose the distribution of the interpolation points $x_i$, one remedy is to cluster them near the end-points of the interval $[a,b]$, 
for example using various kinds of Chebyshev points.

Sowe have to make do with them, and then we need to look for other kinds of interpolants. 
A very popular alternative nowadays is to use splines (piecewise polynomials), which have become a standard tool for many kinds of interpolation and approximation algorithms, 
and for geometric modeling. However, it has been known for a long time that the use of rational functions can also lead to much better approximations than ordinary polynomials. 
In fact, both, polynomial and rational interpolation, can exhibit exponential convergence when approximating analytic functions.
      In ``classical`` rational interpolation, one chooses some $M$ and $N$  such that $M+N=n$ and fits a rational function of the form $\frac{p_M}{q_N}$ to the values $f(x_i)$, 
where $p_M$ and $q_N$ are polynomials of degrees $M$ and $N$ respectively. If $n$ is even, it is typical to set $M+N=\frac{n}{2}$, 
and some authors have reported excellent results. The main drawback, though, is that there is no control over the occurrence of poles in the interval of interpolation.

      \citet{book19} suggested that it might be possible to avoid poles by using rational functions of higher degree. 
They considered algorithms which fit rational functions whose numerator and denominator degrees can both be as high as $n$. 
This is a convenient class of rational interpolants because every such interpolant can be written in so-called barycentric form
\[
 r(x)=\dfrac{\sum\limits_{i=0}^n \dfrac{\lambda_i}{x-x_i}f(x_i)}{\sum\limits_{i=0}^n \dfrac{\lambda_i}{x-x_i}f(x_i)}
\]
for some real values $\lambda_i$. Thus it suffices to choose the weights $\lambda_i$ in order to specify $r$, 
and the idea is to search for weights which give interpolants $r$ that have no poles and preferably good approximation properties. 
Various approach is described in \citet{book30}, in particular, one can choose $\lambda_i=(-1)^i,\quad i=0,1,2,\dots,n$.

\CCLsubsection{Some Other Applications}

PA is widely used for the construction of solitons and other localized solutions of nonlinear problems, even in connection with the appeared term "padeon" \citep{book46, book47}.
         As a simple model, we consider the nonlinear BVP
\begin{eqnarray}
y''-y+2y^3=0,\label{eq3-37}\\
y(0)=1,\quad y(\infty)=0,\label{eq3-38}
\end{eqnarray}
that has an exact localized solution
\begin{equation}\label{eq3-39}
y=\cosh^{-1}(x)
\end{equation}              
	Quasilinear asymptotics give a solution in the following form:
\begin{equation}\label{eq3-40}
y=Ce^{-x}\left(1-0.25 C^2 e^{-2x}+0.0625 C^4 e^{-4x}+\dots\right),\quad C=\mbox{const.}
\end{equation}
     It is easy to verify that with reconstructing the truncated series (\ref{eq3-40}) 
in the PA and to determine the constant $C$ from the BCs (\ref{eq3-38}) and we arrive at the exact solution (\ref{eq3-39}).

It is also interesting to use the PA to the problems with the phenomenon of ``blow-up``, when the solution goes to infinity at a finite value of the argument. For example, the Cauchy problem
\begin{equation}\label{eq3-41}
\dfrac{{\mbox{d}} x}{{\mbox{d}} t}=\alpha x+ \ve x^2,\quad x(0)=1,\quad 0< \ve \ll \alpha \ll 1,
\end{equation}
has the exact solution
\begin{equation}\label{eq3-42}
x(t)=\dfrac{\alpha\exp(\alpha t)}{\alpha + \ve -\ve \exp(\alpha t)},
\end{equation}
which tends to infinity for $t\rightarrow \ln [(\alpha+\ve)/\ve]$.

Regular asymptotic expansion
\[
x(t)\sim \exp(\alpha t)-\ve \alpha^{-1}\exp(\alpha t)[1-\exp(\alpha t)]+\dots 
\]
can not describe this phenomenon, but the use of the PA gives the exact solution (\ref{eq3-42}).

	PA allows to expand the scope of the known approximate methods. For example, 
in the method of harmonic balance the representation of the solution of a rational function of the type
\[
 x(t)=
\dfrac{
\sum\limits_{n=0}^N 
\left\{A_n\cos[(2n+1)\omega t]+B_n\sin[(2n+1)\omega t]\right\}
}
{1+\sum\limits_{n=0}^N \left\{C_m\cos(2m\omega t)+D_m\sin(2m\omega t)\right\}}
\]
substantially increases the accuracy of approximation \citep{book37, book59}.
PA can be used effectively to solve ill-posed problems.
This could include reconstruction of functions in the presence of noise \citep{book34, book35}, 
various problems of dehomogenization (i.e., determining the components of a composite material on its homogenized characteristics) \citep{book24}, etc. 
We must also mention 2D PA \citep{book82}.


\CCLsection{Matching of Limiting Asymptotic Expansions}
\CCLsubsection{Method of Asymptotically Equivalent Functions for Inversion of Laplace Transform}

This method was originally proposed by Slepian and Yakovlev for the treatment of integral transformations. 
Here is a description of this method, following \citet{book73}.
Suppose that the Laplace transform of a function of a real variable $f(t)$ is:
\[
 F(s)=\int\limits_0^\infty f(t)e^{-st}\;{\mbox{d}}s.
\]

To obtain an approximate expression for the inverse transform, it is necessary to clarify the behavior of the transform the vicinity of the points $s=0$ and $s=\infty$ 
and determine the nature and location of its singular points are on the exact boundary of the regularity or near it. 
Then the transform $F(s)$ replaced by the function $F_0(s)$, allowing the exact inversion and satisfying the following conditions:
\begin{enumerate}
\item Functions $F_0(s)$ and $F(s)$ are asymptotically equivalent at $s\rightarrow \infty$ and $s\rightarrow 0$, i.e.
\[
F_0(s)\sim F(s)\quad\mbox{at}\quad s\rightarrow 0 \quad\mbox{and}\quad s\rightarrow\infty.				
\]
\item Singular points of the functions $F_0(s)$ and $F(s)$, located on the exact boundary of the regularity, coincide.
\end{enumerate}
The free parameters of the function $F_0(s)$ are chosen in such a way that they satisfy the conditions of the approximate approximation of $F(s)$ 
in the sense of minimum relative error for all real values $s \geq 0$:
\begin{equation}\label{eq4-1}
\mbox{min}\left\{\mbox{max}\left|\dfrac{F_0(s, \alpha_1,\alpha_2,\dots,\alpha_k)}{F(s)}-1\right|\right\}.
\end{equation}
     Condition (\ref{eq4-1}) is achieved by variation of free the parameters $\alpha_i$. Often the implementation of equalities
\[
 \int\limits_0^\infty F_0(s)\;{\mbox{d}}s=\int\limits_0^\infty F(s)\;{\mbox{d}}s
\]
 or $F_0'\sim F_0'$   at $s\rightarrow 0$ leads to a rather precise fulfillment of the requirements (\ref{eq4-1}).
        Constructed in such a way the function is called asymptotically equivalent function (AEF).

Here is an example of constructing AEF. Find the inverse transform if the Laplace transform is the modified Bessel function \citep[Sect.9]{book1}:
\begin{equation}\label{eq4-2}
 K_0(s)=-\ln(s/2)I_0(s)+\sum\limits_{k=0}^\infty\dfrac{s^{2k}}{2^{2k}(k!)^2}\Psi(k+1)
\end{equation}
where $\Psi(z)$  is the psi (digamma) function \citep[Sect.6]{book1}.

For pure imaginary values of the argument $s (s=iy; \; 0<|y|<\infty )$ function $K_0(s)$ has no singular points.  
Consequently, we can restrict the study of its behavior for $s\rightarrow 0$ and $s\rightarrow\infty$. 
The corresponding asymptotic expressions are \citep[Sect.9]{book1}:
\begin{equation}\label{eq4-3}
\begin{array}{llllll}
 K_0(s)&=&-\left[\ln \dfrac{s}{2}+\gamma\right]+O(s),\quad&s\rightarrow 0,\\\\
 K_0(s)&=&\sqrt{\dfrac{\pi}{2s}}e^{-s}\left[1+O\left(\dfrac{1}{s}\right)\right],\quad &s\rightarrow \infty,&
\end{array}
\end{equation}
where $\gamma$ is the Euler's constant ($\gamma = 1,781\dots$) (note the typo in the first formula (\ref{eq4-3}) in \citet{book73}).

The analyzed Laplace transform has a branch point of the logarithmic type, branch point of an algebraic type, and an essential singularity. 
These singular points need to be stored in the structure of the zero approximation. The most simple way to obtain such a structure, 
combining two asymptotic representations (\ref{eq4-3}) so that they are mutually do not distort each other and contain free parameters, 
which could be disposed of in the future. As a result, we arrive at the zero approximation
\begin{equation}\label{eq4-4}
F_0(s)=e^{-s}\left[\ln\dfrac{s+\alpha}{s}+\sqrt{\dfrac{\pi}{2}}\dfrac{1}{\sqrt{s+\beta}}\right],
\end{equation}
where $\alpha$ and $\beta$ are the free parameters.

 It is easy to see that expression (\ref{eq4-4}) has the correct asymptotic behavior $s\rightarrow\infty$. 
The free parameters are determined from the condition of coincidence of the asymptotics of the functions $K_0(s)$ and $F_0(s)$  for $s\rightarrow 0$ and the equality of integrals
\[
 \int\limits_0^\infty F_0(s)\;{\mbox{d}}s=\int\limits_0^\infty K_0(s)\;{\mbox{d}}s.
\]

     As a result of calculations one obtains a system of transcendental equations
\[
 \begin{array}{rcl}
  \ln \alpha+\sqrt{\dfrac{\pi}{2\beta}}&=&\ln 2-\gamma;\\\\
\ln\alpha-e^\alpha \mbox{Ei}(-\alpha)+\gamma+\dfrac{\pi}{\sqrt{2}}e^\beta\left[1-\mbox{erf}\left(\sqrt{2}\right)\right]&=&\dfrac{\pi}{2},
 \end{array}
\]
where $\mbox{Ei}(\dots)$ is the the sine integral \citep[Sect.5]{book1}, 
$\mbox{erf}(\dots)$ is the the error function \citep[Sect.7]{book1} (note typo in these formulas in \citet{book73}).

Solving the prescription system numerically, one finds  $\alpha = 0.3192$, $\beta=0.9927$.
 
 Then the approximate inverse transform can be written as follows:
\begin{equation}\label{eq4-5}
f_0(t)=\left\{\dfrac{1-\exp[-\alpha(t-1)]}{t-1}+\dfrac{\exp[-\beta(t-1)]}{\sqrt{2(t-1)}}\right\}H(t-1).
\end{equation}
           The exact expression for the function $f(t)$ is:
\begin{equation}\label{eq4-6}
f(t)=\dfrac{1}{\sqrt{t^2-1}}H(t-1)
\end{equation} 

Comparison of exact (\ref{eq4-6}) (solid line) and approximate (\ref{eq4-5}) (dotted line with circles) inversions is shown in Figure 7. 
As it can be seen, a satisfactory result is obtained even in the zero approximation.
\begin{figure}[!ht]
  	\epsfig{figure=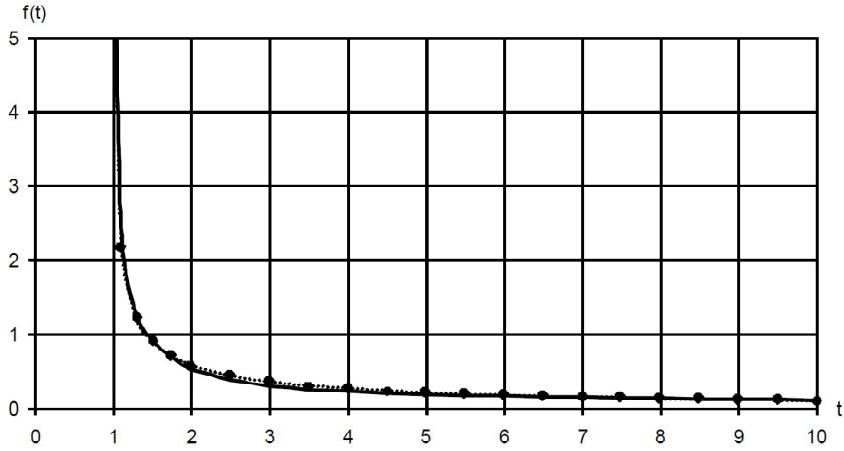,scale=0.22}
	\vspace*{-0.5cm}
	\caption{Comparison of the exact Laplace transform inversion with the treatment by the method of AEFs.}
	\label{Fig7}
\end{figure}
Analogously, one can construct AEFs for inverse sine and cosine Fourier transforms, Hankel and other integral transforms.
 
\CCLsubsection{Two-point Pad\'{e} Approximants}

The analysis of numerous examples confirms: usually implemented a sort of "complementarity principle": 
if for $\ve\rightarrow 0$ one can construct a physically meaningful asymptotics, there is a nontrivial asymptotics and $\ve\rightarrow\infty$. 
The most difficult in terms of the asymptotic approach is the intermediate case of $\ve\sim 1$. 
In this domain numerical methods typically work well, however, if the task is to investigate the solution depending on the parameter  $\ve$, 
then it is inconvenient to use different solutions in different areas. 
Construction of a unified solution on the basis of limiting asymptotics is not a trivial task, which can be summarized as follows: 
we know the behavior of functions in zones I and III (Figure 8), 
we need to construct it in the zone II. For this purpose one can use a two-point Pad\'{e} approximants (TPPA). We give the definition following \citet{book13}. Let
\begin{eqnarray}
F(\ve)=\sum\limits_{i=0}^\infty c_i\ve^i&\mbox{at}&\ve\rightarrow 0,\label{eq4-7}\\\nonumber\\
F(\ve)=\sum\limits_{i=0}^\infty c_i\ve^{-i}&\mbox{at}&\ve\rightarrow \infty.\label{eq4-8}
\end{eqnarray} 
\begin{figure}[!ht]
  	\epsfig{figure=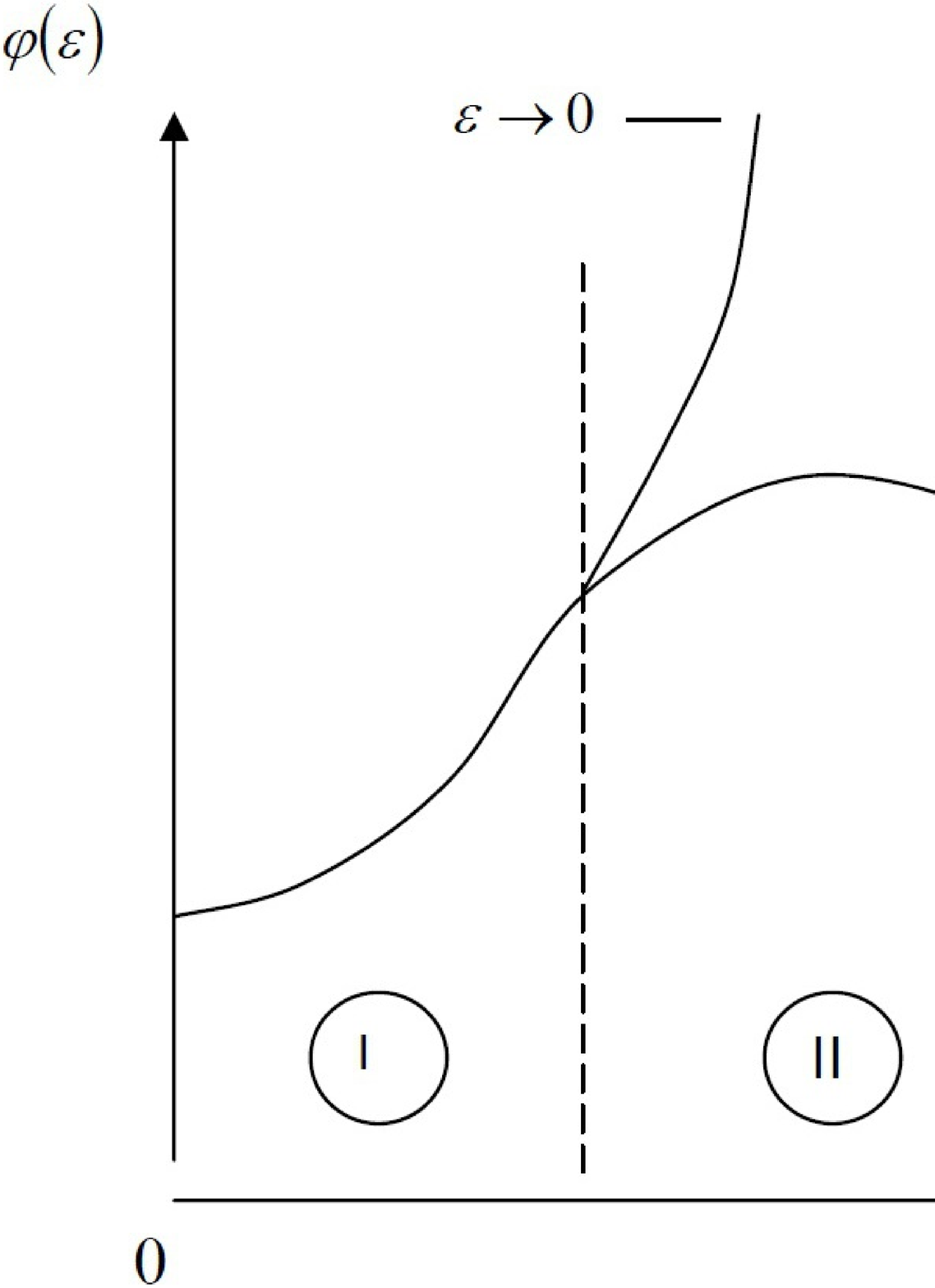,scale=0.23}
	\vspace*{-0.5cm}
	\caption{Matching of asymptotic solutions.}
	\label{Fig8}
\end{figure}

Its TPPA is a rational function of the form 
\[
 f_{[n/m]}(\ve)=\dfrac{a_0+a_1\ve +\dots + a_n\ve^n}{1+b_1\ve+\dots + b_m\ve^m}
\]
with $k\leq m+n-1$ coefficients which are determined from the condition 
\[
 \left(
1+b_1\ve +\dots + b_m\ve^m
\right)
\left(
c_0+c_1\ve + c_2\ve^2+\dots \right)=a_0+a_1\ve +\dots + a_n\ve^n
\]
and the remaining $m+n-k$ coefficients of a similar condition for $\ve^{-1}$.

As an example of TPPA using for matching of limiting asymptotics, consider the solution of the van der Pol equation:
\[
 \ddot x+\ve \dot x\left(x^2-1\right)+x=0.
\]

Asymptotic expressions of the oscillation period for small and large values of $\ve$ are \citep{book38}:
\begin{eqnarray}
T=2\pi\left(1+\dfrac{\ve^2}{16}-\dfrac{5\ve^4}{3072}\right)&\mbox{at}&\ve\rightarrow 0,\label{eq4-9}\\\nonumber\\
T=\ve\left(3-2\ln 2\right)&\mbox{at}&\ve\rightarrow \infty.\label{eq4-10}
\end{eqnarray} 

       For constructing TPPA we use the four conditions at $\ve \rightarrow 0$  and the two conditions at $\ve \rightarrow\infty$, then 
\begin{equation}\label{eq4-11}
T(\ve )=\dfrac{a_0+a_1\ve+a_2\ve^2+a_3\ve^3}{1+b_1\ve+b_2\ve^2},
\end{equation} 
where\\\\
 $a_0=2\pi;$ $a_1\dfrac{\pi^2(3-2\ln 2)}{4(3-2\ln 2)^2-\pi^2};$ $a_2=\dfrac{\pi(3-2\ln 2)^2}{2\left[4 (3-2\ln 2)^2-\pi^2\right]};$\\\\
$a_3=\dfrac{\pi^2(3-2\ln 2)}{16\left[4 (3-2\ln 2)^2-\pi^2\right]};$ $b_1=\dfrac{\pi(3-2\ln 2)}{2\left[4 (3-2\ln 2)^2-\pi^2\right]};$\\\\
$b_2=\dfrac{\pi^2}{16\left[4 (3-2\ln 2)^2-\pi^2\right]}.$\\\\
Table 4 shows the results of the comparison of numerical values of the period, given in \citet{book2}, with the results calculated by formula (\ref{eq4-11}).

  \begin{table}[h!]
  \begin{center}
  $
  \begin{array}{|c|c|c|}
  \hline
  \ve	&T\;\mbox{numerical}		&T\;\mbox{numerical}	\\\hline
  1				&6.66	&6.61	\\\hline
  2				&7.63	&7.37	\\\hline
  3				&8.86	&8.40	\\\hline
  4				&10.20	&9.55	\\\hline
  5				&11.61	&10.81	\\\hline
  6				&13.06	&12.15	\\\hline
  7				&14.54	&13.54	\\\hline
  8				&16.04	&14.96	\\\hline
  9				&17.55	&16.42	\\\hline
  10				&19.08	&17.89	\\\hline
  20				&34.68	&33.30	\\\hline
  30				&50.54	&49.13	\\\hline
  40				&66.50	&65.10	\\\hline
  50				&82.51	&81.14	\\\hline
  60				&98.54	&97.20	\\\hline
  70				&114.60	&113.29	\\\hline
  80				&130.67	&129.40	\\\hline
  90				&146.75	&145.49	\\\hline
  100				&162.84	&161.61	\\\hline
  \end{array}
  $
   \end{center}
  \caption{Comparison of numerical results and calculations using the TPPA.}
  \label{tab4}
  \end{table}

Now we construct inverse Laplace transform with the TPPA. Let the original function is as follows:
\begin{equation}\label{eq4-12}
f(t)=(1+t^2)^{-0.5}.
\end{equation} 
	Asymptotics of this function looks like:
\[
 f(t)\cong
\left\{
\begin{array}{llllll}
1-0.5t^2+\dots &\mbox{at}&t\rightarrow 0,\\
t^{-1}+\dots   &\mbox{at}&t\rightarrow \infty.
\end{array}
\right.
\]
	
TPPA in this case can be written as:
\begin{equation}\label{eq4-13}
f(t)=\dfrac{1+0.5t}{1+0.5t+0.5t^2}.
\end{equation} 
\begin{figure}[!ht]
  	\epsfig{figure=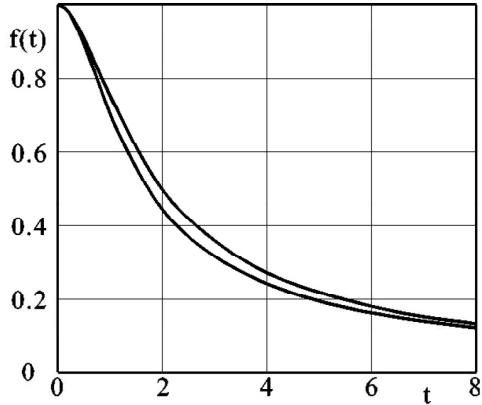,scale=0.20}
	\vspace*{-0.5cm}
	\caption{Exact and approximate Laplace transform inversion.}
	\label{Fig9}
\end{figure}
	Numerical results are shown in Figure 9. An approximate inversion (\ref{eq4-13}) (upper curve) agrees well with the original (\ref{eq4-12}) (lower curve) for all values of the argument.

Other example on the effective use of the TPPA see \citet{book3, book5, book11, book32, book86} .

\CCLsubsection{Other Methods of AEFs Constructiong}

     Unfortunately, the situations where both asymptotic limits have the form of power expansions are rarely encountered in practice, 
so we have to resort to other methods of AEFs constructing.  Consider, for example, the BVP
\begin{equation}\label{eq4-14}
\ve y_{xx}-xy=\ve y,\quad y(0)=1,\quad y(\infty)=0, \ve\quad \ll 1.
\end{equation} 
Solution for small values of $x$ can be written as follows: 
\begin{equation}\label{eq4-15}
 y=1-a\xi+\dfrac{1}{6}\xi^3+O(\xi^4)
\end{equation}
where $\xi=x\ve^{-\frac{1}{3}}$, $a$ is the arbitrary constant.      
                                                              
       The solution for large values of $x$ is constructed using the WKB method \citep{book63}
\begin{equation}\label{eq4-16}
y=b\xi^{-\frac{1}{4}}\exp\left(-\dfrac{2}{3}\xi^{\frac{3}{2}}\right)\left[1-\dfrac{5}{48}\xi^{-\frac{3}{2}}+O\left(\xi^{-3}\right)\right]
\end{equation}
where  $b$ is an arbitrary constant.

         Now we match these asymptotics. Because of the exponential in Eq. (\ref{eq4-16}) using TPPA in the original form is not possible. 
Therefore, we construct AEF, based on the following considerations: 
for large values of the variable $\xi$ exponent from Eq. (\ref{eq4-16}) is taken into account in its original form, 
and for small values of the variable $\xi$ it is expanded in a Maclaurin series. 
Constructed in this way AEF has the form
\begin{equation}\label{eq4-17}
 y_a=\dfrac{1-a\xi+\dfrac{2}{3}\xi^{\frac{3}{2}}-\dfrac{2}{3}\xi^{\frac{5}{2}}+\dfrac{32}{5}a\xi^4}{1+\dfrac{32}{5}\dfrac{a}{b}\xi^{\frac{17}{4}}}\exp\left(-\dfrac{2}{3}\xi^{\frac{3}{2}}\right)
\end{equation}
 The coefficients $a$ and $b$ in Eq. (\ref{eq4-17}) still remain uncertain. 
For calculation of these constants one can use some integral relations, for example, obtained from Eq. (\ref{eq4-17}) 
by multiplying them with the weighting functions $1,x,x^2,\dots$ and further integration over the interval $\left[0,\infty\right)$. In the end, such values of the constants are found:
\begin{equation}\label{eq4-18}
 a=\dfrac{\sqrt[3]{3}\;\Gamma\left(\dfrac{2}{3}\right)}{\Gamma\left(\dfrac{2}{3}\right)},\quad b=\dfrac{\sqrt[3]{9}\;\Gamma\left(\dfrac{2}{3}\right)}{2\sqrt{\pi}}
\end{equation}

       Numerical calculations show that the formula (\ref{eq4-17})  with constants (\ref{eq4-18})  
approximates the desired solution in the whole interval $\left[0,\infty\right)$ with an error not exceeding $1.5\%$.

When choosing the constants one can use other methods, then a lot depends on the skill of the researcher. 
Of course, it is necessary to ensure the correct qualitative behavior of AEFs, avoiding, for example, do not correspond to the problem of zeros of the denominator. 
To do this, one can vary the number of terms in the asymptotics and the numerator and denominator constructed uniformly suitable solutions.
In general form the method of rational AEF can be described as follows \citep{book52}. Let us assume that function $f(z)$ has the following asymptotics:
\begin{equation}\label{eq4-19}
f(z)=F(z)\quad \mbox{at}\quad z\rightarrow\infty,
\end{equation}
and 
\begin{equation}\label{eq4-20}
f(z)=\sum\limits_{i=0}^\infty c_i z^i\quad \mbox{at}\quad z\rightarrow 0.
\end{equation}
    Then the AEF can be produced from the Eqs. (\ref{eq4-21}), (\ref{eq4-22}) as follows:
\begin{equation}\label{eq4-21}
f(z)\approx\dfrac{\sum\limits_{i=0}^m\alpha_i(z) z^i}{\sum\limits_{i=0}^n\beta_i(z) z^i} \quad \mbox{at}\quad z\rightarrow 0.
\end{equation}
where $\alpha_i$, $\beta_i$ are considered not as constants but as some functions of $z$. 
Functions $\alpha_i(z)$  and $\beta_i(z)$ are chosen in such a way that: 
\begin{enumerate}
 \item the expansion of  AEF (\ref{eq4-21}) in powers of $z$ for $z\rightarrow 0$  matches the perturbation expansion (\ref{eq4-20});
 \item the asymptotic behavior of AEF (\ref{eq4-21}) for $z\rightarrow \infty$ coincides with the function $F(z)$ (\ref{eq4-19}).
\end{enumerate}

In the construction of AEFs a priori qualitative information is very important. 
For example, if from any considerations it is known that the unknown function is close to the power, you can use the method of Sommerfeld \citep{book42}. 
Its essence is to replace a segment of the power series
\begin{equation}\label{eq4-22}
f(x)=1+a_1x+a_2x^2+\dots\;,
\end{equation}
to the function
\begin{equation}\label{eq4-23}
f(x)\approx (1+Ax)^\mu.
\end{equation}
Expanding expression (\ref{eq4-23}) in a Maclaurin series and comparing coefficients of this expansion with (\ref{eq4-22}), one obtains
\[
 A=\dfrac{a_1^2-2a_2}{a_1};\quad \mu=\dfrac{a_1^2}{a_1^2-2a_2}
\]

     Numerical approaches also can be used for the construction of AEFs. 
For example, in paper by \citet{book41} a computational technique for matching limiting asymptotics is described.

      Sometimes it is possible to construct so called composite equations, which can be treated as ``asymptotically equivalent equations''.  
Let us emphasize, that the composite equations, due to \citet{book81}, can be obtained in result of synthesis of the limiting cases. 
The principal idea of the method of the composite equations can be formulated in the following way \citep[p.195]{book81}: 
\begin{enumerate}
\item Identify the terms in the differential equations whose neglect in the straightforward approximation is responsible for the nonuniformity.
\item Approximate those terms insofar as possible while retaining their essential character in the region of nonuniformity.
\end{enumerate}

     Let's dwell on the terminology. Here we use the term ``asymptotically equivalent function''. 
Other terms (``reduced method of matched asymptotic expansions`` \citep{book42}, ``quasifractional approximants'' \citep{book23}, 
 ``mimic function'' \citep[p.181--243]{book33} also used. 

\CCLsubsection{Example: Schr\"odinger Equation}

For the Schr\"odinger equation (\ref{eq60})  with boundary conditions (\ref{eq61}) previously we obtained a solution for the exponent, little different from the two (\ref{eq68}). 
In \citep{book20} the following asymptotic solutions for $N\rightarrow\infty$ is obtained: 
\begin{equation}\label{eq4-24}
E_0(N)=\dfrac{\pi^2}{4}(2N)^{-\frac{2}{N+1}}\Gamma\left(\dfrac{N}{N+1}\right)^2.
\end{equation}
     Using (\ref{eq68}) and (\ref{eq4-24}), we construct AEF
\begin{equation}\label{eq4-25}
E_0(N)\sim \dfrac{\pi+\Gamma\left(\dfrac{N}{N+1}\right)^2}{4(2N+\alpha)^{\frac{2}{N+1}}}.
\end{equation}
where $\alpha=\pi^2\Gamma(1.25)-2\approx 6.946$. 
      Numerical results are presented in Table 5. It is evident that formula (\ref{eq4-25}) gives good results for all the values of $N$.
  \begin{table}[h]
  \begin{center}
  $
  \begin{array}{|c|c|c|c|}
  \hline
  N	&E_0\;\mbox{numerical}		&Eq. (\ref{eq4-25})	& \mbox{Error,\% }	\\
        &\mbox{(Boettcher and Bender,}  &			&			\\
        &\mbox{(1990)}			&			&			\\\hline
  1	&1.0000				&1.0			&0			\\\hline
  2	&1.0604				&0.9974			&5.9364			\\\hline
  4	&1.2258				&1.17446		&4.1882			\\\hline
  10	&1.5605				&1.5398			&1.33			\\\hline
  50	&1.1052				&2.1035			&0.079			\\\hline
  200	&2.3379				&2.3376			&0.006			\\\hline
  500	&2.4058				&2.4058			&\approx 0		\\\hline
 1500	&2.4431				&2.4431			&\approx 0		\\\hline
 3500	&2.4558				&2.45558		&\approx 0		\\\hline
  \end{array}
  $
   \end{center}
  \caption{Comparison of numerical and analytical results of the energy levels for the Schrödinger equation.}
  \label{tab5}
  \end{table}

\CCLsubsection{Example: AEFs in the Theory of Composites}

 Now let us consider an application of the method of AEFs for the calculation of the effective heat conductivity of an infinite regular array of perfectly conducting spheres, 
embedded in a matrix with unit conductivity. \citet{book68} have obtained the following expansion for the effective conductivity $\left\langle k\right\rangle$:
\begin{equation}\label{eq4-26}
\left\langle k\right\rangle=1-
\dfrac{3c}
{
-1+c+a_1c^{\frac{10}{3}}\dfrac{1+a_2c^{\frac{11}{3}}}{1-a_3c^{\frac{7}{3}}}+a_4c^{\frac{14}{3}}+a_5c^{6}+a_6c^{\frac{22}{3}}+O\left(c^{\frac{25}{3}}\right)
},
\end{equation}
where $c$ is the volume fracture of inclusions.
      Here we consider three types of space arrangement of spheres, namely, the simple cubic (SC), 
body centered cubic (BCC) and face centered cubic (FCC) arrays. The constants $a_i$ for these arrays are given in Table 6.
  \begin{table}[h]
  \begin{center}
  $
  \begin{array}{|c|c|c|c|c|c|c|}
  \hline
			&a_1		&a_2		&a_3		&a_4		&a_5		&a_6	\\\hline
  \mbox{SC array}	&1.305		&0.231		&0.405		&0.0723		&0.153		&0.0105	\\\hline
  \mbox{BCC array}	&0.129		&-0.413		&0.764		&0.257		&0.0113		&0.00562\\\hline
  \mbox{FCC array}	&0.0753		&0.697		&-07.41		&0.0420		&0.0231		&9.14\cdot 10^{-7}\\\hline
  \end{array}
  $
   \end{center}
  \caption{ The constants $a_1,\dots, a_6$ in Eq. (\ref{eq4-26}).}
  \label{tab6}
  \end{table}

     In the case of perfectly conducting large spheres ($c\rightarrow c_{max}$, where $c_{max}$ is the maximum volume fraction for a sphere) 
the problem can be solved by means of a reasonable physical assumption that the heat flux occurs entirely in the region where spheres are in a near contact. 
Thus, the effective conductivity is determined in the asymptotic form for the flux between two spheres, which is logarithmically singular in the width of a gap, 
justifying the assumption \citep{book56}:
\begin{equation}\label{eq4-27}
\left\langle k\right\rangle=-M_1\ln\chi-M_2+O(\chi^{-1}),
\end{equation}
where $\chi=1-\left(\frac{c}{c_{max}}\right)^{\frac{1}{3}}$ is the dimensionless width of a gap between the neighboring spheres,
$\chi\rightarrow 0$ for $c \rightarrow c_m$, $M_1=0.5c_{max}p$, $p$  is the number of contact points at the surface of a sphere; $M_2$ is a constant, 
depending on the type of space arrangement of spheres. 
The values of $M_1$, $M_2$ and $c_{max}$  for the three types of cubic arrays are given in Table 7.

  \begin{table}[h]
  \begin{center}
  $
  \begin{array}{|c|c|c|c|}
  \hline
			&M_1		&M_2		&c_{max}	\\\hline
  \mbox{SC array}	&\pi/2		&0.7		&\pi/6		\\\hline
  \mbox{BCC array}	&\sqrt{3}\pi /2	&2.4		&\sqrt{3}\pi /8		\\\hline
  \mbox{FCC array}	&0\sqrt{2}\pi	&7.1		&\sqrt{2}\pi/6		\\\hline
  \end{array}
  $
   \end{center}
  \caption{ The constants $M_1$, $M_2$ and $c_{max}$.}
  \label{tab7}
  \end{table}

\begin{figure}[!ht]
  	\epsfig{figure=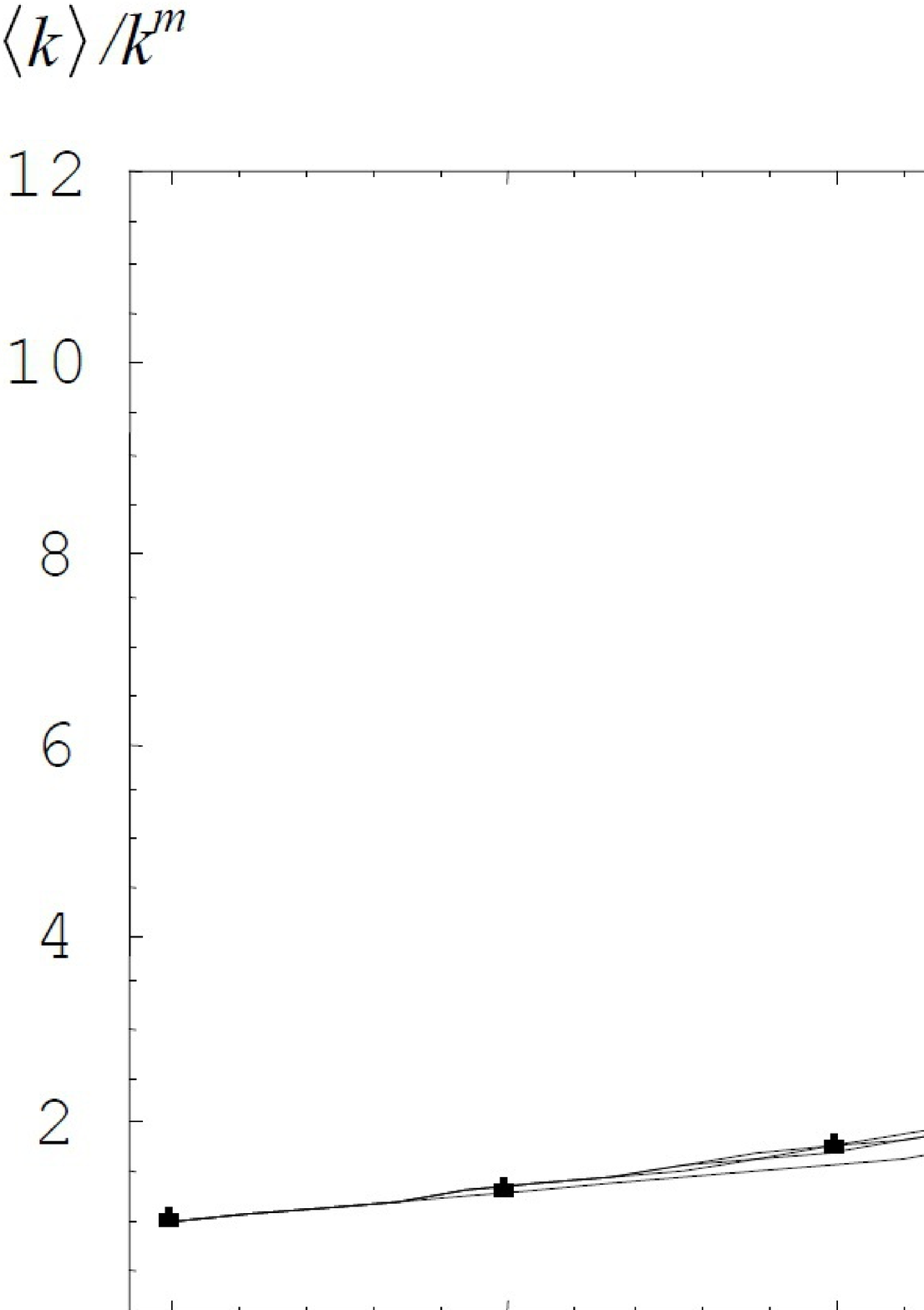,scale=0.15}
	\vspace*{0.3cm}
	\caption{Effective conductivity $\left\langle k\right\rangle / k^m$ of the SC array vs. volume fraction of inclusions $c$.}
	\label{Fig10}
\end{figure}
\begin{figure}[!ht]
  	\epsfig{figure=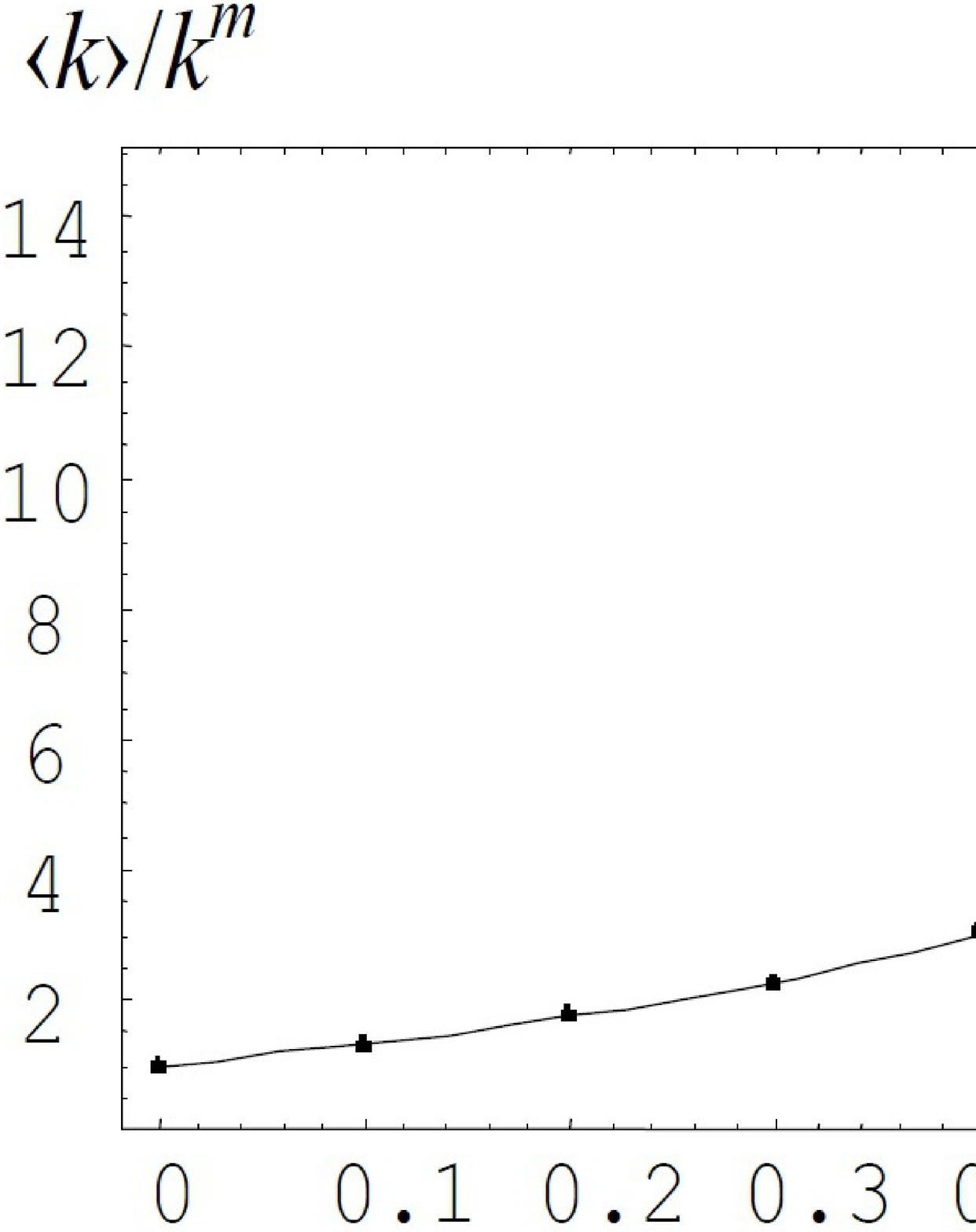,scale=0.20}
	\vspace*{-0.5cm}
	\caption{Effective conductivity $\left\langle k\right\rangle / k^m$ of the BCC array vs. volume fraction of inclusions $c$.}
	\label{Fig11}
\end{figure}

\begin{figure}[!ht]
  	\epsfig{figure=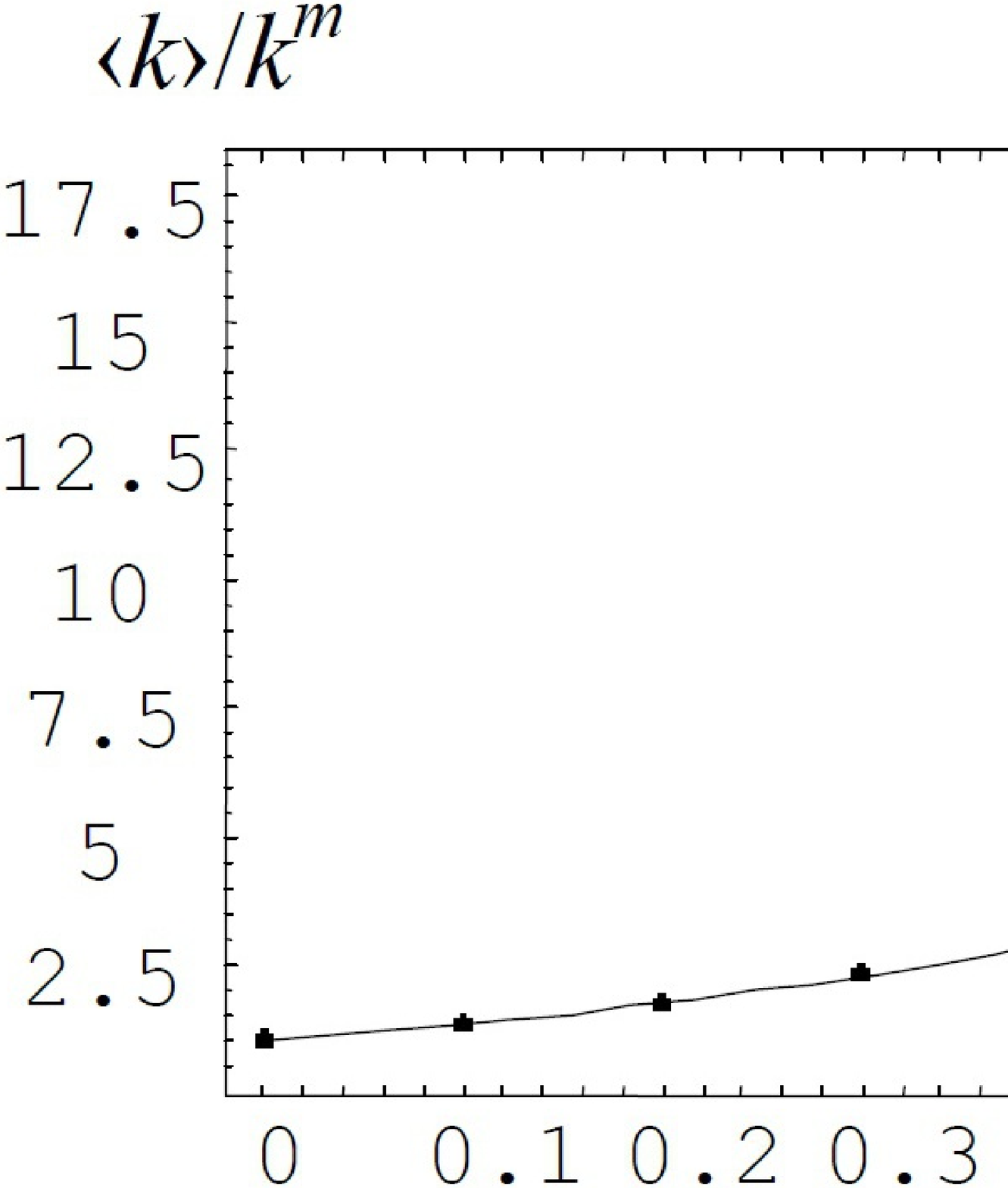,scale=0.20}
	\vspace*{-0.5cm}
	\caption{Effective conductivity $\left\langle k\right\rangle / k^m$ of the FCC array vs. volume fraction of inclusions $c$.}
	\label{Fig12}
\end{figure}
     On the basis of limiting solutions (\ref{eq4-26}) and (\ref{eq4-27}) we develop the AEF valid for all values of the volume fraction of inclusions 
$c\in \left[0, c_{max}\right]$:
\begin{equation}\label{eq4-28}
\left\langle k\right\rangle=
\dfrac{P_1(c)+P_2c^{\frac{m+1}{3}}+P_3\ln\chi}{Q(c)},
\end{equation}
       Here the functions $P_1(c)$, $Q(c)$ and the constants $P_2$, $P_3$ are determined as follows:\\\\
$
 Q(c)=1-c-a_1c^{\frac{10}{3}},\quad P_1(c)=\sum\limits_{i=0}^{m}\alpha_1 c^{\frac{i}{3}},\quad P_2=0 \quad \mbox{for}\quad n=1,\\\\
P_2=\dfrac{-\left[P_1(c_{max})+Q(c_{max})M_2\right]}{c_{max}^{\frac{m+1}{3}}}\quad\mbox{for }n=2.\\\\
$
     The AEF (\ref{eq4-28}) takes into account  leading terms of expansion (\ref{eq4-26}) and  leading terms of expansion (\ref{eq4-27}). Coefficients  are:\\\\
$
\alpha_0=1,\quad \alpha_3=2-\dfrac{Q(c_{max})M_1}{3c_{max}},\quad \alpha_{10}=\alpha_1-\dfrac{Q(c_{max})M_1}{10c_{max}^{\frac{10}{3}}},\\\\
\alpha_j=-\dfrac{Q(c_{max})M_1}{jc_{max}^{\frac{j}{3}}},\quad j=1,2,\dots,m-1m,,\quad j\neq 3,10.
$

      The increment of $m$ and $n$ leads to the growth of the accuracy of the obtained solution (\ref{eq4-28}). Let us illustrate this dependence in the case of SC array. 
We calculated  $\left\langle k\right\rangle$ for different values of $m$ and $n$. 
In the Figure 10 our analytical results are compared with experimental measurements from \citet{book57} (black dots). 
Details of these data can be found in \cite{book55}. 
Finally, we restrict $m=19$ and $n=2$  for all types of arrays, as they provide a satisfactory agreement with numerical datas 
and a rather simple analytical form of the AEF (\ref{eq4-28}).

      Numerical results for the BCC and the FCC arrays are displayed in Figures 11 and 12 respectively. 
For BBC array the obtained AEF (\ref{eq4-28}) is compared with the experimental results from \citet{book53} and \citet{book54}. 
For FCC array the experimental data are not available, therefore we are comparing with the numerical results obtained by \citet{book54} using the Rayleigh method. 
The agreement between the analytical solution (\ref{eq4-28}) and the numerical results is quite satisfactory.

    \bibliographystyle{plainnat}
    \bibliography{Contribs}

\end{document}